\pdfoutput=1
\documentclass{JHEP3}
\usepackage{amsmath}
\usepackage{graphicx}
\usepackage{mathrsfs}
\usepackage{mathcomp}
\usepackage{enumerate}
\usepackage{cite}
\usepackage[utf8]{inputenc}
\usepackage{xspace}
\usepackage{ifthen}
\usepackage{url}
\usepackage{booktabs}
\usepackage{multirow}
\usepackage{subcaption}

\newcommand{\pythia}{P\protect\scalebox{0.8}{YTHIA}\xspace}

\newcommand{\pytppp}{P\protect\scalebox{0.8}{YTHIA}8\xspace}

\newcommand{\alice}{ALICE\xspace}
\newcommand{\cms}{CMS\xspace}

\newcommand{\angantyr}{Angantyr\xspace}
\newcommand{\rivet}{Rivet\xspace}

\newcommand{\etal}{\textit{et~al.}}
\newcommand{\push}{nudge\xspace}

\newcommand{\pushes}{nudges\xspace}

\renewcommand{\eqref}[1]{eq.~(\ref{#1})\xspace}

\newcommand{\fig}[1]{\ref{#1}}
\newcommand{\figref}[1]{figure~\fig{#1}}

\newcommand{\Figref}[1]{Figure~\fig{#1}}

\newcommand{\citeref}[1]{ref.~\cite{#1}}
\newcommand{\citerefs}[1]{refs.~\cite{#1}}

\newcommand{\sect}[1]{\ref{#1}}
\newcommand{\sectref}[1]{section~\sect{#1}}
\newcommand{\sectrefs}[1]{sections~\sect{#1}}

\newcommand{\appref}[1]{appendix~\sect{#1}}

\def\text{\mathrm}
\def\eg{\emph{e.g.}\xspace}
\def\ie{\emph{i.e.}\xspace}
\def\cf{\emph{c.f.}\xspace}

\def\mrm#1{\mathrm{#1}}

\def\tauT{\ensuremath{\tau_{T}}}
\def\tauH{\ensuremath{\tau_{H}}}
\def\tauS{\ensuremath{\tau_{S}}}

\def\epem{$e^+e^-$}

\def\pp{\ensuremath{\mrm{pp}}\xspace}

\def\pA{\ensuremath{\mrm{p}A}\xspace}

\def\AA{\ensuremath{AA}\xspace}

\def\PbPb{\ensuremath{\mrm{PbPb}}}

\def\kappaLG{\ensuremath{\kappa_{LG}}}
\def\kappaS{\ensuremath{\kappa}} 
\def\etaS{\ensuremath{\eta_{st}}} 
\newcommand{\diff}[1]{\ensuremath{d#1}}
\newcommand{\dtwo}[1]{\ensuremath{d^2#1}}
\newcommand{\dthree}[1]{\ensuremath{d^3#1}}

\def\ColText(#1,#2)[#3]#4{\Text(#1,#2)[#3]{#4}}

\preprint{LU-TP 20-48\\
  MCnet-20-22\\
  arXiv:2002.nnnnn [hep-ph]\\
}

\title{
  Setting the string shoving picture in a new frame.
  \footnote{Work supported in part by the Knut and Alice Wallenberg
foundation, contract number 2017.0036, the Swedish
    Research Council, contracts number 2016-03291, 2016-05996 and
    2017-0034, in part by the European Research Council (ERC) under
    the European Union’s Horizon 2020 research and innovation
    programme, grant agreement No 668679, and in part by the MCnetITN3
    H2020 Marie Curie Initial Training Network, contract 722104.  }}

\author{Christian Bierlich, Smita Chakraborty, Gösta Gustafson, and Leif Lönnblad\\
  Dept.~of Astronomy and Theoretical Physics,
  Sölvegatan 14A, S-223 62 Lund, Sweden\\
  E-mail: \email{christian.bierlich@thep.lu.se},
  \email{smita.chakraborty@thep.lu.se},
  \email{gosta.gustafson@thep.lu.se}, and
  \email{leif.lonnblad@thep.lu.se}}

\abstract{Based on the recent success of the \angantyr model in
  describing multiplicity distributions of the hadronic final state in
  high energy heavy ion collisions, we investigate how far one can go
  with a such a string-based scenario to describe also flow effects
  measured in such collisions.

  For this purpose we improve our previous so-called \textit{shoving}
  model, where strings that are close in space--time tend to repel
  each other in a way that could generate anisotropic flow, and we
  find that this model can indeed generate such flows in \AA\
  collisions. The flow generated is not quite enough to reproduce
  measurements, but we identify some short-comings in the presented
  implementation of the model that, when fixed, could plausibly give a
  more realistic amount of flow. }

\begin{document}

\sloppy

\newpage
\section{Introduction}
\label{sec:intro}

High energy proton-proton and heavy ion (HI) collisions are frequently
analysed assuming very different dynamical mechanisms.  Models based
on string formation or cluster chains and subsequent hadronization,
implemented in event generators like \pythia
\cite{Sjostrand:2014zea,Sjostrand:2006za}, HERWIG
\cite{Bellm:2015jjp,Bahr:2008pv}, or SHERPA
\cite{Bothmann:2019yzt,Gleisberg:2008ta}, have been
very successful in describing particle production in
$\mathrm{e}^+\mathrm{e}^-$ annihilation and pp collisions.  For pp
collisions this picture includes multi-parton subcollisions, and at
high energies the scattered partons are mainly gluons, which form
colour connected gluon chains. This picture corresponds to multiple
pomeron exchange (including pomeron vertices), with a gluon chain
pictured as a cut BFKL pomeron ladder.  It is also consistent with
models based on reggeon theory, \eg\ the models by the groups in
Tel~Aviv \cite{Gotsman:2014pwa}, in Durham \cite{Khoze:2014aca}, or by
Ostapchenko \cite{Ostapchenko:2010vb}. The gluons here also hadronize
forming strings or cluster chains.

On the other hand, many features in HI collisions can be
described assuming the formation of a thermalized quark--gluon plasma (QGP), in
particular collective flow \cite{Broniowski:2008vp,
Schenke:2010rr,Song:2011hk,Gardim:2012yp,Gale:2012rq,Noronha-Hostler:2015uye} and 
enhancement of strange particles \cite{Andronic:2017pug}.
The formation of a hot thermalized plasma is often described as follows: At
high energy the gluons in a nucleus form a ``color glass condensate'' (CGC)
\cite{Kovner:1995ja, Kovner:1995ts}. This state is described by a
classical colour field, where the strength is saturated due to unitarity, with
a ``saturation scale'' $Q_s \sim (1/\alpha_s) (x/x_0)^{-\lambda}$. 
When two nuclei collide the overlapping non-Abelian fields form a
``glasma''; for reviews of the CGC and the glasma see \eg\ \cite{Lappi:2006fp,
 Venugopalan:2008xf, Gelis:2012ri}. The glasma state contains parallel
longitudinal colour-electric and colour-magnetic fields, associated with
induced magnetic charges in the projectile and target remnants. 
These fields also build up a topological charge giving rise to CP-violating
effects \cite{Kharzeev:1998kz, Kharzeev:2001ev}. The glasma is instable, and
in this scenario it turns rapidly into a QGP, which soon
thermalizes \cite{Chen:2015wia}.
This transition is often motivated referring to 
``Nielsen--Olesen instabilities'' in the QCD Fock vacuum \cite{Nielsen:1978rm, Nielsen:1978nk}, or ``Weibel instabilities'' in an electro-magnetic plasma 
\cite{Weibel:1959zz}.

Lately the difference between pp and nuclear collisions has become much less
clear. Both collective 
effects and increased strangeness production have been observed in
high multiplicity pp events (see \eg\ \cite{Khachatryan:2010gv} and
\cite{ALICE:2017jyt}). This has raised the question whether a plasma
can be formed in pp collisions, or alternatively, if these effects in
HI collisions also can be described in a scenario based on strings.

In high multiplicity pp events, the density of strings is quite high. The
width of a stringlike flux tube is estimated to be of the order of
$1/\Lambda_{QCD}$, which implies that the flux tubes will overlap in space.  
Recently we have demonstrated, that in pp collisions both collectivity and
strangeness enhancement
can be explained as consequences of a higher energy density in 
systems of overlapping strings or ``ropes''. This gives both a transverse
pressure, which can cause long range collective flow \cite{Bierlich:2017vhg},
and enhanced strangeness following from the higher energy release in the breakup
of a rope \cite{Bierlich:2014xba}.
At the same time string-based models have successfully described the general
particle distributions in nuclear collisions \cite{Wang:1991hta,
  Bierlich:2018xfw}. In this paper we will show that some  of the collective effects
seen in $AA$ collisions can be
described by the transverse pressure from overlapping strings, in the same way
as in pp collisions, without assuming the 
formation of a thermalized state with high temperature. The strangeness
enhancement in $AA$ expected when nearby strings form ``ropes'', will be
discussed in a later publication. We expect this effect to be similar, but
also further enhanced in $AA$ compared to pp, which was described in
\citeref{Bierlich:2014xba}. 

The two scenarios discussed here are qualitatively different. It is important
to note that the difference
lies not in the initial gluonic states in the 
projectile and the target. In the pp MC models mentioned above, the parton
distributions are fitted to HERA data, and thus include the observed
suppression for $Q^2 < Q_s^2(x)$, as a result of saturation at small $x$.
They also agree with BFKL evolution
including saturation effects, as formulated in the dipole model DIPSY
\cite{Avsar:2006jy, Gustafson:2015ada}, which has the same
evolution with rapidity as the CGC, including the suppression for $q_\perp <
Q_s(x)$, as an effect of Debye screening at high gluon density. In the limit
$N_c \rightarrow \infty$ this follows because the gluon emissions split the
dipoles into gradually smaller and smaller dipoles. For finite $N_c$ the
screening effects are further enhanced by identical colours in neighbouring
dipoles, described by colour reconnection in the initial state 
cascade via the ``swing
mechanism'' in the DIPSY model \cite{Avsar:2006jy}.

The difference lies instead in the assumed behaviour just after the collision.
The particle distribution is (in symmetric collisions) characterized by an
approximately boost invariant central plateau. This feature is quite natural
if the hadrons are produced from boost invariant strings stretching between
the projectile and target remnants. The energy per unit length in a string,
$\diff{E}/\diff{z}$, is given by the string tension $\kappaS \approx 0.9$~GeV/fm.  
The energy in the collision is initially concentrated in the forward-going
remnants. These gradually loose energy, which is stored in the new string
pieces, as the strings become longer. 
The energy density, $\diff{E}/\dthree{x}$, is therefore approximately constant from the
instant of collision until the hadronization (proper) time $\tauH \sim
1.5-2$~fm (see \sectref{sec:lundstring}). From this moment the hadrons move 
out with rapidities approximately equal to the hyperbolic angle $\etaS
\equiv (1/2) \ln[(t+z)/(t-z)]$ corresponding to the place where they were
"born".\footnote{This space--time coordinate is conventionally called $\eta$,
  but should not be mixed up with pseudo-rapidity 
 $= (1/2) \ln[(p+p_z)/(p-p_z)]$. To avoid confusion we will below use the
 notation $\etaS$ for the space--time coordinate.} 
In the center the relation between $\etaS$ and $z$ is given by
\begin{equation}
\Delta z = t \,\Delta \etaS. 
\label{eq:eta-y-relation} 
\end{equation}
For a single string the final state energy density in rapidity is therefore
given by (replacing $t$ by the boost invariant variable $\tau= \sqrt{t^2-z^2}$) 
\begin{equation}
\diff{E}/\diff{y} = \kappaS/\tauH.
\end{equation}

A boost invariant plateau is also expected from a glasma, if the
longitudinal colour-electric and -magnetic fields are stretched out in a
similar way. But the glasma is expected to rapidly transform into a
thermalized plasma, and this plasma will expand longitudinally and dilute
with an energy density falling like $\tau^{-1}$ or faster, before freezeout or
hadronization time \cite{Bjorken:1982qr}. The energy density in the glasma
must therefore correspond 
to the (energy density at freezeout)$\times (\tauH/\tauT)$,
where $\tauT$ is the time for the glasma-plasma transition.

The different energy densities also implies that, in the string picture, the 
features of the vacuum condensate is very important to keep the field together 
in the strings, while the condensate energy is too small to play an important 
role in the plasma scenario, where it is normally neglected; see \eg\ 
\citeref{Chen:2015wia} where this is explicitly stated.
The features of the energy density and
the vacuum condensate are discussed in somewhat more detail in
\sectref{sec:differences}.

If both approaches are able to reproduce presently available data on, in
particular, collective effects and strangeness enhancement, it becomes
important to find observables where they will give different predictions. As an
example, one such observable would be effects of CP violations, where clear
signals are not expected in the string scenario.

We have already presented some preliminary studies in
\citeref{Bierlich:2017vhg} where we used a simplified so-called
\textit{shoving} model to describe how strings in a \pp\ collision
repel each other to create the so-called near-side ridge first found
in \pp by \cms at the LHC \cite{Khachatryan:2010gv}. The simplified
model had a number of short-comings. One was that it only treated
strings that were parallel to the beam direction, using an upper cut
on the transverse momentum of the partons in the string. Another was
that the force between the strings was manifested in terms of many
very soft gluons, which was technically difficult to handle in the
\pytppp hadronization implementation, and also somewhat difficult to
reconcile within the string model as such.\footnote{For a discussion
  about very soft gluons in the Lund model see \eg\
  \cite{Andersson:1998sw}.}

In this work we will present a more advanced model, where all string
pieces in an event can interact, not only in \pp, but also in \AA\ and
basically in any other collision system. Furthermore the problem with
producing too many additional soft gluons is circumvented by applying
the force in terms of tiny transverse momentum \pushes given directly
to the produced hadrons in the hadronization. Even though the new
model does not produce extra soft gluons, it still has some problem
dealing with soft gluons already present, stemming from the
perturbative phase in \pytppp, that complicates the string motion,
and currently this restricts the amount of shoving that can be
achieved. 

The layout of this paper is as follows. In \sectref{sec:differences}
we first further elaborate the motivation behind trying to treat
high density \AA\ collisions in terms of interacting strings, rather
than as a hot thermalized plasma of free partons. There we also
discuss in some detail what is known about the structure of a QCD
string using the analogy with a vortex lines in a superconductor. In
\sectref{sec:lundstring} we then look at how the QCD string is
described in the Lund string fragmentation model. Our new shoving
model is presented in \sectref{sec:thepicture} giving some details of
how one finds a Lorentz frame where any two string pieces always lie
in parallel planes, and how we there can discretize the shoving into
tiny transverse \pushes between them which are then applied to the
hadrons produced. We study the behaviour of the new model in
\sectrefs{sec:simple-aa} and \ref{sec:results}. First we apply it to a
toy model for the geometry of the initial partonic state of an \AA\
collision to show that we can qualitatively describe some of the flow
effects found there. Then we apply the new model to a more realistic
initial state generated by the \angantyr model \cite{Bierlich:2018xfw}
for \AA\ collisions in \pytppp, and find that the quantitative
description if flow effects \AA\ collisions is still lacking. Finally
in \sectref{sec:conclusions} we present our conclusion and present
some ideas for improvement of our new model that may achieve an
improved description of experimental measurements.

\section{Differences between a thermalized and a non-thermalized scenario}
\label{sec:differences}

\subsection{Initial energy density and possible plasma transition}
\label{sec:density}

As mentioned in the introduction, one of the most popular scenarios  for treating heavy ion collisions, involves 
early time dynamics calculated as quasi-classical gluon fields immediately after
the collision (a ``glasma''), followed by a thermalized plasma phase treated with (3+1D) hydrodynamics
\cite{Schenke:2010rr}. 
The initial glasma state contains longitudinal colour-electric and colour-magnetic fields associated with induced magnetic charges in the forward-moving nucleus remnants. Longitudinal electric or magnetic fields are boost invariant, have no longitudinal momentum and a constant energy per unit length. When the longitudinal glasma fields become extended, transverse fields are generted, which keep the flux contained in tubes. In \citeref{Chen:2015wia} it is shown that the energy density is proportional to $(1-\mathcal{O}((Q\tau)^2))$, where $Q$ is a typical scale related to the saturation scale $Q_s$.
Such nearly constant initial fields have essential similarities with the relativistic strings in the Lund model. The glasma fields are, however, assumed to rapidly turn into a thermalized QGP, which expands longitudinally, with an energy density falling like $\tau^{-1}$ (or faster) \cite{Bjorken:1982qr}. To give the observed energy density after freezeout, the energy density before thermalization must be correspondingly higher.

This picture can be contrasted to a string scenario, where the longitudinal energy 
density $dE/dz$ is constant until hadronization. As discussed in the introduction, 
if we look in the rest frame of the hadrons
produced by a string piece with hyperbolic angle $\Delta \etaS$, then at proper time $\tau$
this piece corresponds to a distance $\Delta z=\tau \Delta \etaS$ in the normal coordinate
space.

When the string hadronizes the average density in rapidity $y$ is the same as
in $\etaS$, giving for a single string an energy density $\diff{E}/\diff{y} \approx \diff{E}/\diff{\etaS} = \kappaS \cdot \tauH$ with $\kappaS \approx 0.9$ GeV/fm and 
the hadronization time $\tauH = 1.5-2$ fm (see \sectref{sec:lundstring}). 
Minijets will enhance the energy density $\diff{E}/\diff{y}$ in the lab frame, but also
here the initial density is limited.

The energy density and the time for thermalization in the glasma state
are difficult to estimate theoretically. They are frequently present
in lattice units or arbitrary units (\eg\ in
\citeref{Chen:2015wia}) or they can be adjusted to fit experimental
data as \textit{e.g.} in \citeref{Schenke:2012wb}.  Tuning to a
hydrodynamical plasma calculation for RHIC data at 200 GeV and $b=9$
fm (centrality $\approx 40$ \%) \cite{Schenke:2012wb}, finds an
average density
$\diff{E}/\tau \,\diff{y} \approx \diff{E}/\diff{z} \approx 300$
GeV/fm. The experimental result for this centrality is
$\diff{N_{ch}}/\diff{y} \approx 150$ \cite{Bearden:2001qq,
  Jipa:2004ny}. Adding neutrals and assuming
an average transverse mass $\sim$ 0.5 GeV, this gives
$dE/dy \approx 110$ GeV, which agrees with \eqref{eq:eta-y-relation}
if the thermalization time is about 0.35 fm. The energy density would
then be $\diff{E}/\dthree{x} \sim 8\, \textrm{GeV/fm}^3$.  A
corresponding calculation for the string scenario, where the strings
hadronize at $\tau = 1.5 - 2$ fm, would instead give the density 1.7
$\textrm{GeV/fm}^3$.

In a central PbPb collision at LHC energy, the charged particle density 
$\diff{N_{ch}}/\diff{y}$ is of
the order 2000 (\cf\ the data from \alice shown in \figref{fig:Pb-multiplicity}).
Including neutrals gives approximately 3000, and again assuming an
average transverse mass $\sim 0.5$~GeV and
a transverse area $\sim 150 \,\,\mathrm{fm}^2$, this corresponds to an
energy density $\diff{E}/\diff{y}\, \dtwo{x_\perp} \sim 10 \,\,\mathrm{GeV/fm}^2$. 
In the string scenario, 
\eqref{eq:eta-y-relation} gives an initial energy density in coordinate space
$\diff{E}/\dthree{x} \approx 5-6\,\, \mathrm{GeV/fm}^3$. 
The energy density in the glasma is also calculated for LHC energy by Schenke \textit{et al.} in 
\citeref{Schenke:2012hg}. The energy density $dE/\tau dy \approx dE/dz$, obtained at $\tau = 0.1$ fm, is about 20,000 GeV/fm, or $dE/d^3 x \approx 500\, \textrm{GeV/fm}^3$. 
At this time the transverse and longitudinal fields have become approximately equally strong, and
this density also roughly matches the density needed to reproduce the experimental data from \alice, for a thermalization at $\tau = 0.1$ fm. We note that this energy density is almost a factor 100 larger than what is needed in the string scenario. (We also note that the energy density obtained in
\citeref{Schenke:2012hg} is not approximately constant for small times, but instead falls rapidly from approximately 50,000 GeV/fm at very early times to about 20,000 GeV/fm at $\tau = 0.1$ fm.)

The conclusion from these examples is that the initial energy density needed in the glasma is one or two orders of magnitude larger than in the string scenario. This also implies that, while the energy density in the vacuum condensate can safely neglected in the glasma, it is important in the string scenario.

\subsection{The vacuum condensate and dual QCD}
\label{sec:condensate}

As discussed above, the initial energy density is relatively low
in the non-thermal scenario, and the properties of the vacuum condensate is
important for the formation of colour-electric flux tubes
(strings).
The QCD vacuum was early studied in several papers by N.K. Nielsen,
H.B. Nielsen and P. Olesen. In \citeref{Nielsen:1978rm} it was shown
that the QCD Fock vacuum (with zero fields) is
instable. If a small homogeneous colour-magnetic field is added, it will
grow exponentially. However, as shown in the accompanying paper
\cite{Nielsen:1978nk}, if the externally added field is in a given direction
colour space, it will induce fields in the orthogonal directions in
colour space. (For simplicity the authors studied SU(2) Yang--Mills theory.)
Higher orders in this induced field then develop a Higgs potential, analogous
to the potential describing the condensate in a superconductor. This implies
that the exponential growth will be stopped, and the
initial Fock vacuum will fall into a non-perturbative ground state with
negative energy. 
\footnote{It is often argued that the Nielsen--Olesen instability contributes
to a rapid transition from a glasma to a plasma. Nielsen and Olesen showed that
the Fock vacuum is a local maximum, and thus unstable. However, it is not
obvious to us 
how this effect motivates a fast transition when a perturbation is added to the
(stable) real vacuum.}

In a normal superconductor a magnetic field is
compressed in vortex lines or flux tubes, and magnetic charges would be
confined. The pure Yang--Mills theory
is symmetric under exchange of electric and magnetic fields. Quarks with
colour-electric charge are confined, and the
Copenhagen vacuum is also known to have non-trivial
vortex solutions of electric type (see \eg\ \citerefs{Nielsen:1979xu,
Ambjorn:1978ff}). Thus the QCD vacuum behaves as a
\textit{``dual superconductor''}, with exchanged roles for the electric and the
magnetic fields. For a review of
the ``Copenhagen vacuum'', see \eg\ \cite{Olesen:1980nz}.

It was further demonstrated by 't~Hooft that the ground state in a pure SU(3) 
Yang--Mills theory can have two different phases, with either colour-electric
or colour-magnetic flux tubes, but not both \cite{tHooft:1979rtg}. In the first 
phase colour-electric charge is confined, and extended colour-magnetic
strings are not possible, while the opposite is true in the second phase.
The fact that quarks are confined obviously shows that the QCD vacuum is of 
the first kind.

Although the pure Yang--Mills theory is symmetric under exchange of electric and magnetic fields, Maxwell's equations for electromagnetism are asymmetric, due to the absence of magnetic charges. Exchanging electric and magnetic fields is obtained by 
replacing the field $F^{\mu\nu}$ by the dual field tensor $\widetilde{F}^{\mu\nu}\equiv \epsilon^{\mu\nu\kappa\lambda} F_{\kappa\lambda}$. The electric and magnetic currents would then be given by
\begin{equation}
j^\mu_{el}= \partial_\nu F^{\nu\mu},\,\,\, 
j^\mu_{magn}= \partial_\nu \widetilde{F}^{\nu\mu}.
\label{eq:currents} 
\end{equation}
Expressing the field $F^{\mu\nu}$ as derivatives of a vector potential 
$A^\mu$ implies, however, that the magnetic current is identically zero.
As a consequence magnetic charges can only be introduced by adding extra degrees of freedom. 
Dirac restored the symmetry between electric and magnetic charges by adding a ``string term'', an antisymmetric tensor 
$G_{\mu\nu}$ satisfying $\partial_\nu G^{\nu\mu} = j^\mu_{magn}$. The electromagnetic field is then given by 
\begin{eqnarray}
F_{\mu\nu} = \partial_\mu A_\nu - \partial_\nu A_\mu - \widetilde{G}_{\mu\nu},\nonumber \\
\widetilde{F}_{\mu\nu} = \widetilde{\partial_\mu A_\nu} - \widetilde{\partial_\nu A_\mu} +G_{\mu\nu}.
\label{eq:Diracstring}
\end{eqnarray}
A consistency constraint is then a quantization of electric ($e$) and magnetic ($g$) charges: $e\cdot g = 2\pi n$ with $n$ an integer.

The features of a vacuum behaving as a dual superconductor was early
discussed in a series of papers by Baker, Ball, and Zachariasen; for a review
see \citeref{Baker:1991bc}. In a dual superconductor a colour-electric flux 
tube has to be kept together by a colour-magnetic current. Instead of expressing 
the extra degrees of freedom in terms of Dirac's string term, Baker \textit{et al.} treated $F^{\mu\nu}$ and $\widetilde{F}^{\mu\nu}$ as independent fields. 
(In the non-trivial vacuum condensate they are related by a non-local magnetic permeability.)
Higher order corrections then induce a Higgs potential in the $\widetilde{F}$ field, analogous to the induced field in the Nielsen--Olesen instability.

The fields $F^{\mu\nu}$ and $\widetilde{F}^{\mu\nu}$ studied by Baker \textit{et al.} 
are non-Abelian. A problem is here that, although a bound $q\bar{q}$ pair must be a colour
singlet, and the energy density has to be gauge invariant, this is not the case
for the colour-electric and -magnetic fields.'t~Hooft has, however, shown 
that the essential confining features can be 
described by an Abelian subgroup $U(1)^2$ to SU(3) \cite{tHooft:1981bkw}.
More recent studies of dual QCD are based on this ``Abelian dominance''. 
This is also supported by lattice calculations performed in the maximal Abelian
gauge, which exhibit a confining phase related to the condensation of magnetic
monopoles \cite{Carmona:2001ja,Cea:2001an}. 
For overviews of Abelian projections (or Abelian gauge fixing) see also \citerefs{Ripka:2003vv, DiGiacomo:2017blx}.

\subsection{A single QCD flux tube in equilibrium}
\label{sec:singlestring}

A normal superconductor can be described by the Landau--Ginzberg (LG) equations,
see \eg\ \citeref{ref:deGennes} and \appref{sec:vortex}.
Here the vacuum condensate is formed by Cooper pairs and influenced by a Higgs
potential. In the interior of the superconductor a magnetic field is kept
inside flux tubes or vortex lines by an electric current,
 and magnetic monopoles would be confined. The
flux in such a vortex line (and the charge of a possible monopole) is quantized
in multiples of $2\pi/q$, where $q=2e$ is the charge of a Cooper pair.

A vortex line or a flux tube is characterized
by two fundamental scales: the penetration depth $\lambda$ and the coherence
length $\xi$. These scales are the inverse of, respectively, the mass attained by
the gauge boson and the mass of the Higgs particle.
At the boundary between a superconducting and a normal phase, if
$\lambda \gg \xi$ the parameter $\lambda$ determines how far the magnetic field 
penetrates into the condensate (from which it is expelled by the
Meissner effect). Similarly if $\xi \gg \lambda$, then $\xi$ determines the rate
at which the condensate goes to zero at the boundary.
When $\xi > \lambda$ (or more exactly $ \xi>\sqrt{2}\,\lambda$), both the
condensate and the field are suppressed over a range $\xi-\lambda$. This is a
type I superconductor, and it implies that the surface provides a
positive contribution to the energy. In equilibrium the surface then tends to
be as small as possible. If in contrast $\lambda$ is larger than $\xi$
(type II superconductor), the condensate and the field can coexist over
a range $\sim\lambda-\xi$, and the surface provides a negative contribution to
the energy, favouring a large surface.
(The properties of a classical superconductor are discussed in somewhat more
detail in \appref{sec:vortex}.)

As discussed above, the QCD vacuum has important similarities with a
superconductor.
There are, however, also important differences between a non-Abelian
flux in QCD and an Abelian field in a non-relativistic superconductor.
The infrared problems in QCD contribute to the difficulties to estimate the
properties of a QCD flux tube. The total energy, given by the string tension,
is fairly well determined from the spectrum of quarkonium bound states,
and approximately equal to 1~GeV/fm. The total flux and also the width of 
the tube are, however, less well known. One problem is that it is
not obvious how much of the string tension is in the field, how much
is in destroying the condensate, and how much is in the current keeping the flux
together inside the tube. Another problem is that, although
the flux is given by the running strong coupling, the scale is not 
well specified. Also for a fixed total flux, the energy stored in the 
linear colour-electric field depends on the (not well known) width of the tube.  
The field energy is $\sim A\, E^2 \sim \Phi^2/A$, where $\Phi$ is the total flux
and $A$ the transverse area of the tube.
We here briefly discuss three approaches to estimate the properties
of a QCD flux tube:

\paragraph{i) The bag model:}

The simplest model for a QCD flux tube is the MIT bag model \cite{Johnson:1975zp}. 
Here the vacuum condensate is destroyed within a radius $R$ around the center
of the tube. Inside the tube there is a homogeneous longitudinal colour-electric
field $E=\Phi/(\pi R^2)$, where $\Phi$ is the total flux. The energy per unit
length of the tube is then
\begin{equation}
\kappaS=\pi R^2\left[(\Phi/\pi R^2)^2/2 +B \right],
\label{eq:bagmodel}
\end{equation}
where the bag constant $B$ is (minus) the energy density in the condensate.

Equilibrium is obtained by minimizing $\kappaS$, which gives $\pi R^2=\Phi^2/2B$
and $\kappaS=2\pi R^2 B$. Here half of the energy is in the field, and half 
comes from destroying the vacuum condensate inside the tube. With 
$\kappaS\approx 1$~GeV/fm and $B\approx
(145\mathrm{~MeV})^4$ \cite{Johnson:1975zp}, we find $R\approx 1.7$~fm, or 
$\sqrt{\langle \rho^2 \rangle} \approx 1.2$~fm. (Here $\rho$ is the radial 
distance in cylinder coordinates.)

\paragraph{ii)  Dual QCD:}

In the approach by Baker \textit{et al.} the fields $\mathbf{D}$ and 
$\mathbf{H}$ are derivatives of a dual potential $C_\mu$.
In the vacuum condensate the magnetic permeability $\mu$ is non-local, and 
the fields $\mathbf{E}=\mu\mathbf{D}$ and $\mathbf{B}=\mu \mathbf{H}$ are 
treated as independent fields (note that $\mu\epsilon=1$) related to a tensor
field $\widetilde{F}^{\mu\nu}$. This tensor field interacts via a Higgs 
potential, forming a vacuum condensate which confines the colour-electric 
field $\mathbf{D}$. 
In \citeref{Baker:1991bc} the LG parameter $\kappaLG = 
\lambda/\xi$ is estimated to $\approx 0.75$, when fitting the model to the string 
tension and the energy in the vacuum condensate. This is quite close to 
the borderline, $1/\sqrt{2}$, between type I and type II superconductors. 
It implies that the surface energy is small, and the authors conclude that 
the behaviour is not far from a flux tube in the bag model.
The width $\sqrt{\langle \rho^2 \rangle}$ is estimated to 0.95~fm.
A more recent review over dual QCD is found in \citeref{Kondo:2014sta}, which also
discusses results from lattice calculations.

\paragraph{iii) Lattice QCD:}

Lattice calculations are natural tools to solve infrared problems, and several groups
have presented studies of flux tubes using different methods; for some recent analyses
see \citerefs{Cea:2014uja, Baker:2018mhw, Baker:2019gsi, Nishino:2019bzb, Shibata:2019ghh, Shibata:2012ae, Battelli:2019lkz}. 
Such analyses are, however, not without problems. One problem is
that the strings have to be rather short, in order to have a good resolution. A good
resolution is important for the determination of the behaviour for small $\rho$.
Besides the colour-electric field, the features of the flux tube also depends on the
properties of the vacuum condensate.

Some studies find that the vacuum acts like type II superconductor, \eg\
\citerefs{Baker:2018mhw,Battelli:2019lkz}, but a majority of recent studies conclude instead that it should be of type I, \eg\ \citerefs{Cea:2014uja, Baker:2019gsi, 
Shibata:2019ghh}. In most analyses the electric field is fitted
using Clem's ansatz \cite{ref:Clem} for the condensate (the order 
parameter) $\psi \propto f \,e^{-i\phi}$, where $f=\rho/\sqrt{\rho^2+\xi_v^2}$. Here $\rho$ 
is the radius in cylinder 
coordinates, and the parameter $\xi_v$ is varied to minimize
the string energy. As discussed in more detail in \appref{sec:vortex}, this ansatz 
satisfies one of LG's two equations. 
Minimizing the string tension then gives the electric field
\begin{equation}
E \propto K_0(\sqrt{\rho^2 + \xi_v^2}/\lambda).
\label{eq:ClemE}
\end{equation}
Here $K_0$ is a modified Bessel function, and the scale parameter
$\lambda$ equals the penetration depth in the LG equations.
The parameter $\xi_v$ is 
tuned to fit the shape of the lattice result for small $\rho$-values, where 
it suppresses the logarithmic singularity in $K_0$. The result is related, 
but not equal to the coherence length in the superconductor.
An important problem is, however, that the ansatz in \eqref{eq:ClemE} is expected to 
work well for type II superconductors, where it gives an approximate solution 
also to the second of LG's two equations. 
It does, however, not work for type I superconductors,
where the second LG equation can be quite badly violated.
This is a problem for those studies, which find fits which correspond to 
$\kappaLG$-values in the type I-region, where the fitted values of $\lambda$ and $\xi$
no longer represent the initial parameters in the LG equations.
This problem is also discussed further in \appref{sec:vortex}. 
 
Another problem is how to translate the width of the field from lattice units to
the physical scale in fm.
This problem is discussed in a review by Sommer in \citeref{Sommer:2014mea}. In earlier studies it was common to adjust the energy in
the colour-electric field, $\int 2\pi \rho\, \diff{\rho} E^2/2$, to the string tension,
known to be $\approx 1$~GeV/fm. Also, as noted above, it is uncertain how much of 
the string tension is due to the field,
how much is due to breaking the condensate, and how much is due to the current which keeps the
flux together.
In the bag 
model or the dual QCD estimate mentioned above, about half of the string energy 
is in the field, and half is due to the condensate. Adjusting the field to
represent the full string tension is therefore likely to overestimate the field 
strength and thus underestimate its radius.

Another way to determine the scale, used in the lattice analyses mentioned 
above, is to study the transition of the $q\bar{q}$ potential
from small to larger separations $r$, \eg\ using a fit of the type
$V(r)=A/r +Br +C$, see \eg\ \citerefs{Edwards:1997xf, Necco:2001xg}.
The parameters can here be adjusted \eg\ to a phenomenological fit to 
quarkonium spectra. This method is, however, also uncertain.
In the review by Sommer, mentioned above, it is concluded that ``the connection of 
the phenomenological potentials to the static potential $V(r)$ has never been 
truely quantitative''.

Although there are uncertainties in the determination of the LG parameter using Clem's approximation, it does give a good description of the \emph{shape} of the
longitudinal field obtained in the lattice calculations. As an example we show in
\figref{fig:e-profile} the result in \citeref{Cea:2014uja}. We here note that
the profile is also quite well represented by a Gaussian distribution. Due to the 
uncertainties in determining the value of $\kappaLG$ and the physical scale, we 
will below approximate the field by a Gaussian determined by two tunable parameters,
which are related to the radius of the field and to the fraction of the 
string tension related to the colour-electric field energy.

\FIGURE[t]{
  \includegraphics[width=0.8\textwidth]{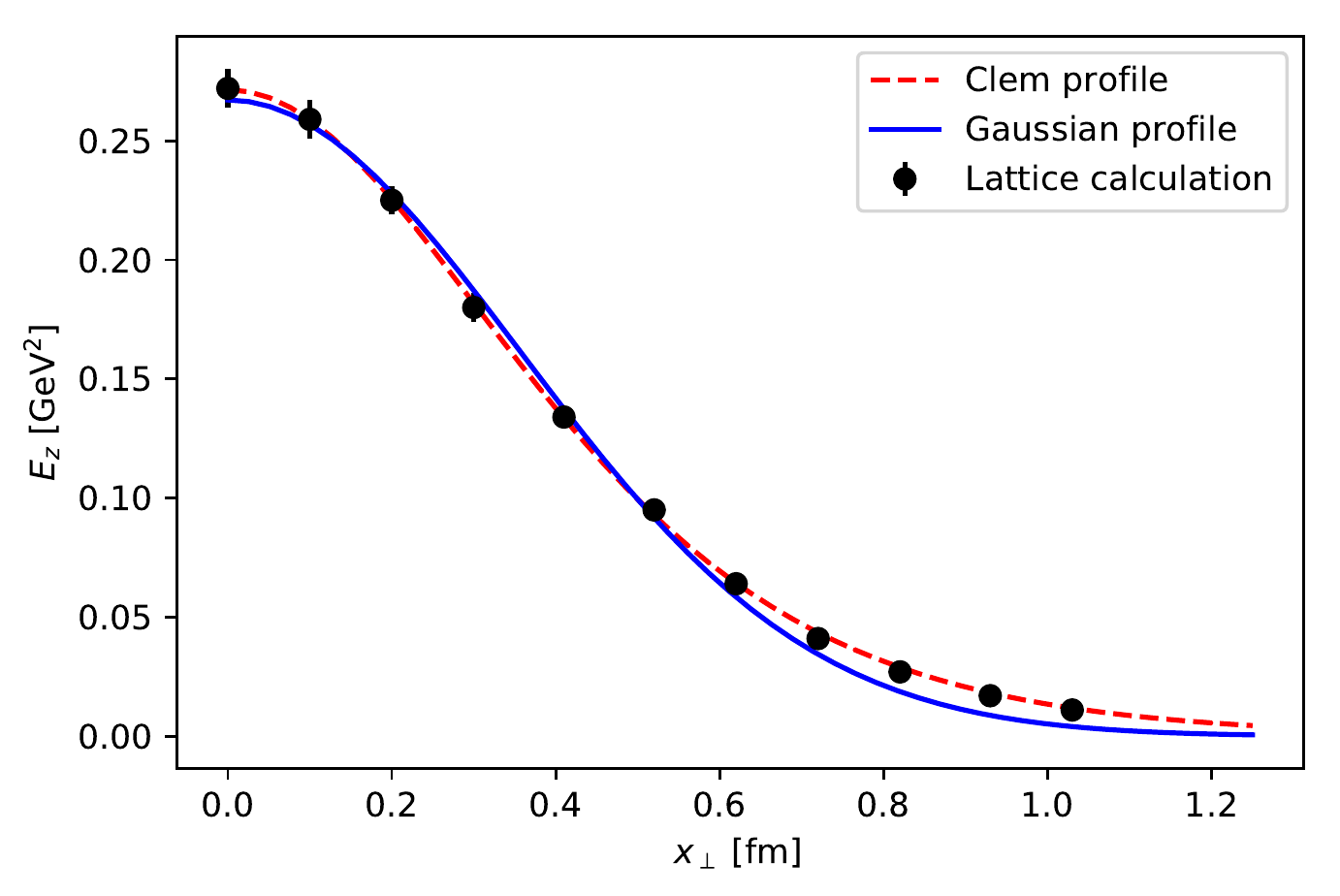}
  \caption{\label{fig:e-profile}Profile of the electric field from the lattice 
  calculation in \citeref{Cea:2014uja} compared to the fit by Clem \cite{ref:Clem} and a Gaussian
  distribution.}
  }

\section{The Lund string hadronization}
\label{sec:lundstring}

In the Lund string hadronization model, described in
\citeref{Andersson:1983ia}, it is assumed that the dynamics of a single flux
tube, and its breakup via quark pair production, is insensitive to the width
of the tube. It is then approximated by an infinitely thin ``massless 
relativistic string''. Essential features of the Lund 
hadronization model are first $q\bar{q}$ pair production via the Schwinger 
mechanism, and secondly the interpretation of gluons as transverse 
excitations on the string. This last assumption implies that the 
hadronization model is infrared safe and insensitive to the addition 
of extra soft or collinear gluons.

\paragraph{i) Breakup of a straight string}

We here first describe the breakup of a straight flux tube or string between a quark and
an antiquark. For simplicity we limit ourself to the situation with a single hadron 
species, neglecting also transverse momenta. For a straight string 
stretched between a quark and an anti-quark, the breakup to a state with 
$n$ hadrons is in the model given by the expression:
\begin{equation}
   \label{eq:lu-had}
    \diff{\mathcal{P}} \propto \prod_{i=1}^n \left[N\dtwo{p_i} \delta(p^2_i - m^2)\right]
    \delta^{(2)} \!\left(\sum p_i - P_{\mathrm{tot}}\right)\exp\left(-bA\right).
\end{equation}
Here $p_i$ and $P_{\mathrm{tot}}$ are two-dimensional vectors.
The expression is a product of a phase space factor, where the parameter $N$
expresses the ratio between the phase space for $n$ and $n-1$ particles, and
the exponent of the imaginary part of the string action,
$bA$. Here $b$ is a parameter and $A$ the space--time area covered by the
string before breakup (in units of the string tension $\kappaS$). 
This decay law can be implemented as an iterative process, where each
successive hadron takes a fraction $z$ of the remaining light-cone momentum
($p^\pm = E \pm p_z$) along the positive or negative light-cone 
respectively. The values of these momentum fractions are then given by the
distribution
\begin{equation}
    \label{eq:lu-frag}
    f(z) = N\frac{(1-z)^a}{z}\exp(-bm^2/z).
\end{equation}
Here $a$ is related to the parameters $N$ and $b$ in \eqref{eq:lu-had} by
normalization. (In practice $a$ and $b$ are determined from
  experiments, and $N$ is then determined by the normalization constraint.)

In applications it is also necessary to account for different quark and hadron
species, and for quark transverse momenta. The result using Schwinger's 
formalism for electron production in a homogenous electric field gives an 
extra factor $\exp(-\pi(\mu^2+p_\perp^2)/\kappaS)$, where $\mu$ and $p_\perp$
are the mass and transverse momentum for the quark and anti-quark in the 
produced pair. For details see \eg\ \citeref{Andersson:1983ia}. 

The result in \eqref{eq:lu-frag} is in principle valid for strings stretched between
partons produced in a single space--time point, and moving apart as illustrated 
in the space--time diagram in fig.~\ref{fig:fan}.
\FIGURE[t]{
  \centering
    \includegraphics[width=0.6\textwidth,bb=200 430 600 690,clip]{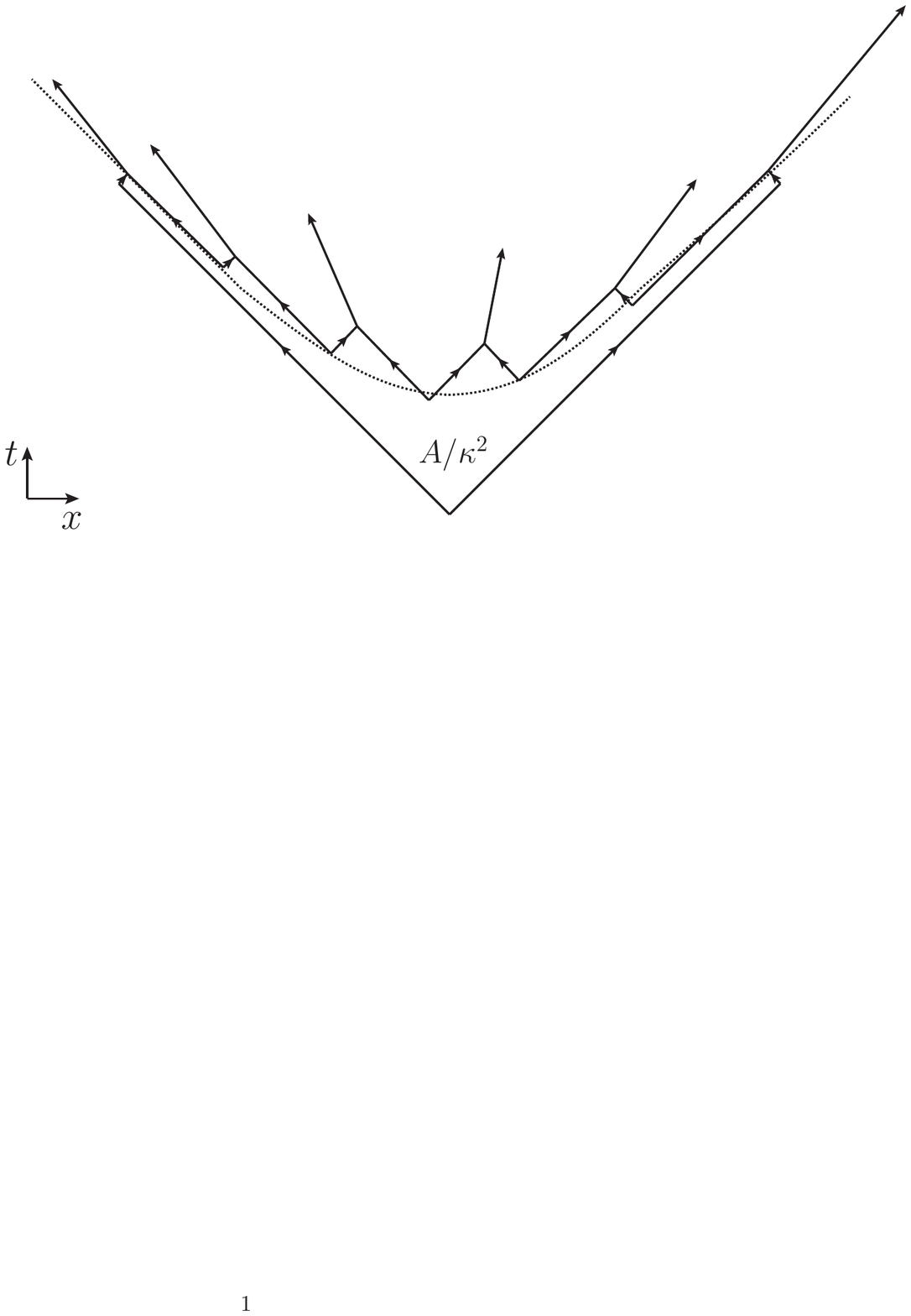}    

  \caption{\label{fig:fan}Breakup of a string between a quark and an
    anti-quark in a $x-t$ diagram. New $q\bar{q}$ pairs are produced
    around a hyperbola, and combine to the outgoing hadrons. The
    original $q$ and $\bar{q}$ move along light-like trajectories. The area
    enclosed by the quark lines is the coherence area $A$ in
    \eqref{eq:lu-had}, in units of the string tension $\kappaS$.
    The notion of the "hadronization time" is not well defined. It could be
    the time when the new $q\bar{q}$ pairs are produced, or when they meet for
    the first  time to form a hadron (or something in between).
}
}
The expression in \eqref{eq:lu-had} is boost invariant, and the hadrons are
produced around a hyperbola in space--time. A Lorentz boost in the
$x$-direction will  
expand the figure in the $(t+x)$ direction and compress it in the $(t-x)$
direction  
(or \textit{vice versa}). Thus the breakups will be lying along the same
hyperbola, and  
low momentum particles in a specific frame will always be the first to be
produced in that special frame.  

The typical proper time for the breakup points is given by
\begin{equation}
    \label{eq:prod-time}
    \langle \tau^2 \rangle = \frac{1 + a}{b\kappaS^2}.
\end{equation}
This does, however, not necessarily correspond to the "hadronization time",
which might 
also be defined as the time when a quark and an anti-quark meet for the first
time to form a hadron. 

With parameters $a$ and $b$
determined by tuning to data from \epem\ annihilation at LEP, and $\kappaS$
equal to 0.9-1~GeV/fm, \eqref{eq:prod-time} gives a typical breakup time 1.5~fm, while the 
average time for the hadron formation is 2~fm. The typical hadronization time can therefore 
be estimated to 1.5-2~fm. This is important to keep in mind, as this value sets
an upper limit on the time available for strings to interact. 

\paragraph{ii) Hadronization of gluons}

An essential component in Lund hadronization is the treatment of gluons.
Here it is assumed that the width of the flux tube can be neglected, and
that its dynamics can be approximated by an infinitely thin ``massless 
relativistic string''.\footnote{For the dynamics of such a Nambu--Goto string, 
see \eg\ \citeref{Artru:1979ye}.} A quark at the endpoint of
the string, carrying energy and momentum, moves along a straight line,
affected by a constant force given by the string tension 
$\kappaS$, reducing (or increasing) its momentum. A gluon is treated as a ``kink'' 
on the string, carrying energy and momentum and also moving along a 
straight line with the speed of light. A gluon carries both colour 
and anti-colour, and the string can be stretched from a quark, via a set 
of colour-ordered gluons, to an anti-quark (or alternatively as a closed 
gluon loop). Thus a gluon is pulled by two string 
pieces, and retarded by the force $2\kappaS$. When it has lost its energy, 
the momentum-carrying kink is split in two corners, which move with the 
speed of light but carry no momentum.
\FIGURE[t,h]{
  \centering
    \includegraphics[width=0.85\textwidth]{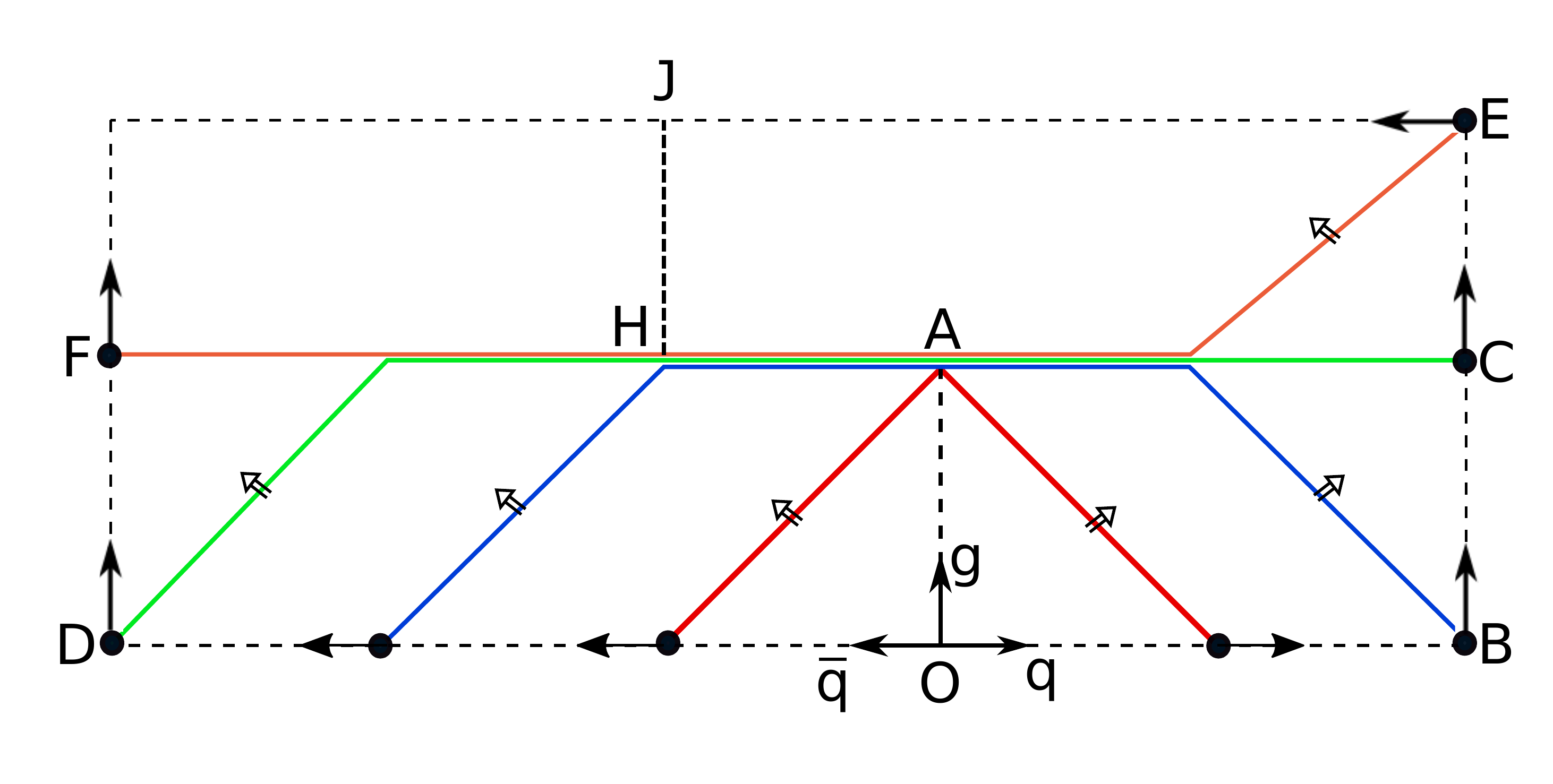}    
   \caption{A $q-g-\bar{q}$ system moving out from 
   a single point $O$ with energies 2, 2, and 3 units of energy respectively.
   The position of the string is shown at four consecutive times, marked 
   by red, blue, green, and purple colour respectively.}
   \label{fig:threejet}
}

A simple example is shown in \figref{fig:threejet}.
It shows a quark, a gluon, and an anti-quark moving
out from a single point, called $O$ in the figure. They move outwards at right
angles with energies 2, 2, and 3 units respectively. 
After 1 unit of time (equal to $\kappaS^{-1}$ energy units) the gluon has
arrived at point $A$ and lost all its energy. The gluon is then replaced by
two corners connected by a straight section. The quark has lost its energy 
in point $B$. In the
rest frame of the attached string piece it now turns back,
gaining momentum in the opposite direction. In the figure frame it is turning
$90^\circ$, and after meeting the string
corner at $C$, it is instead pulled back and loosing energy. The anti-quark
turns around in a similar way at point $D$. If the string does not break
up in hadrons, the string evolution will be reversed. The kinks meet at 
point $H$, and the whole system
collapses to a point $J$. (Note that this system is not in its rest frame.)

We note that although the string pieces initially move with a transverse velocity 
$1/\sqrt{2}$, after some time most of the string is at rest (the horizontal string pieces in 
\figref{fig:threejet}.
A soft gluon will soon stop and be replaced by a straight section stretched
as if it were pulled out between the quark and the anti-quark. This implies that
the string hadronization model is infrared safe; a soft gluon will
only cause a minor modification on the string motion. The same is also 
true for a collinear gluon \cite{Andersson:1983ia}.
For a string with several gluons there will also be several new straight string pieces, 
which become more and more aligned with the directions of the endpoints, as described
in \citeref{Sjostrand:1984ic}. Therefore a string stretched over many rapidity 
units, and with several soft gluon kinks, will be pulled out in a way much 
more aligned with the beam axis, before it breaks into hadrons.

\section{The string shoving picture}
\label{sec:thepicture}

\def\sinth{\ensuremath{\sin\frac{\theta}{2}}}
\def\costh{\ensuremath{\cos\frac{\theta}{2}}}
\def\sinph{\ensuremath{\sin\frac{\phi}{2}}}
\def\cosph{\ensuremath{\cos\frac{\phi}{2}}}
\def\sinhe{\ensuremath{\sinh\frac{\eta}{2}}}
\def\coshe{\ensuremath{\cosh\frac{\eta}{2}}}

In the following we will describe the details in our new
implementation of the string shoving model. First we recap the main
idea as applied to two straight parallel string pieces, then we
consider the interaction between two arbitrary string pieces, and
describe a special Lorentz frame in which the shoving can be properly
formulated. Finally we describe how we discretize the shoving into
small fixed-size \textit{\pushes} and how these are ordered in time and
applied to the final state hadrons.

\subsection{Force between two straight and parallel strings}
\label{sec:parallelforce}

The force between two straight and parallel strings was discussed in 
\citeref{Bierlich:2017vhg}, We here shortly reproduce the treatment presented
there. Just after the production of a 
string stretched between a quark and an anti-quark, the colour field is 
necessarily compressed, not only longitudinally but also transversely. 
They then expand transversely with the speed of light until they reach 
the equilibrium radius $R_S\sim$~0.5-1~fm. 

As illustrated in \figref{fig:e-profile}, the electric field $E$ 
obtained in lattice calculations, and fitted to the Clem 
formula, \eqref{eq:ClemE}, is also well approximated by a Gaussian
\begin{equation}
E =N \exp(-\rho^2/2R^2).
\end{equation}
The normalization factor $N$ can be 
determined if the energy in the field (per unit length), given by 
$\int \dtwo{\rho}\, E^2/2$, is adjusted
to a fraction $g$ of the string tension $\kappaS$. This gives 
$N^2=2g\kappaS/(\pi R^2)$. As discussed in \sectref{sec:singlestring}, the
simple bag model would give $g=1/2$. Due to the uncertainties in determining
the properties of the flux tube, we will treat $R$ and $g$ as tunable
parameters.

When the colour-electric fields in two nearby parallel strings 
overlap, the energy per unit length is given by 
$\int \dtwo{\rho}\, (\mathbf{E}_1 +\mathbf{E}_2)^2/2$. For a transverse 
separation $d_\perp$ this gives the interaction energy 
$2\kappaS g\exp(-d_\perp^2/(4R^2))$. Taking the derivative with respect
to $d_\perp$ then gives the force per unit length
\begin{equation}
f(d_\perp)=\frac{g\kappaS d_\perp}{R^2} \exp\left(-\frac{d_\perp^2}{4R^2}\right).
\label{eq:force}
\end{equation}

For a boost invariant system it is convenient to introduce hyperbolic
coordinates
\begin{equation}
\tau = \sqrt{t^2 - z^2}, \,\,\,\,\, \etaS = \ln ((t+z)/\tau).
\label{eq:hyper}
\end{equation}
Near $z=0$ we get $\delta z=t\, 
\delta \etaS$, and the force in \eqref{eq:force} gives $\diff{p_\perp}/\diff{t}\,
\delta z=f(d_\perp)$. 
Boost invariance then gives the two equations
\begin{equation}
\frac{d p_\perp}{\tau \diff{\tau}\, \diff{\etaS}} =
f(d_\perp),\,\,\,\,\frac{\dtwo{d_\perp}}{\diff{\tau^2}} = \frac{f(d_\perp)}{\kappaS}. 
\label{eq:dpt}
\end{equation}

We have here assumed that the flux tubes are oriented in the same
direction, leading to a repulsion. We have also argued in terms of an
Abelian field, which means that if the fields are oppositely oriented
there would be an attenuation of the fields rather than a
repulsion. Since QCD is non-Abelian, the picture is slightly more
complex, but the calculations are still valid. In case of two triplet
fields in opposite directions, we get with probability 8/9 an octet
field, which also leads to a repulsion when compressed.  Only with
probability 1/9 we get a singlet field, and in this case the strings
are assumed to be removed through ``colour reconnection'', as
described in \citeref{Bierlich:2014xba}. Also for strings in other
colour multiplets the string-string interaction is dominantly
repulsive.  This is not in conflict with the Abelian approximation, as
discussed in \sectref{sec:singlestring}.  A $q\bar{q}$ string is a
colour singlet, where the quark is a coherent mixture of red, blue,
and green (with corresponding anti-colours for the
anti-quark). Similarly the endpoint of an octet string has a coherent
combination of the 8 different colour charges. 

\subsection{String motion and the parallel frame}
\label{sec:parallel-frame}

For a string piece between two (massless) partons, the motion and
expansion of the sting is very simple in the rest frame of the two
partons. If the partons have momenta along the $x$-axis, the position
of the string ends are simply $x_\pm(t)=(t; \pm t, 0, 0)$, where we
note that the ends move by the speed of light irrespective of the
momentum of the partons.

\FIGURE[t]{
  \centering
    \includegraphics[width=0.99\textwidth,bb=140 440 525 630,clip]{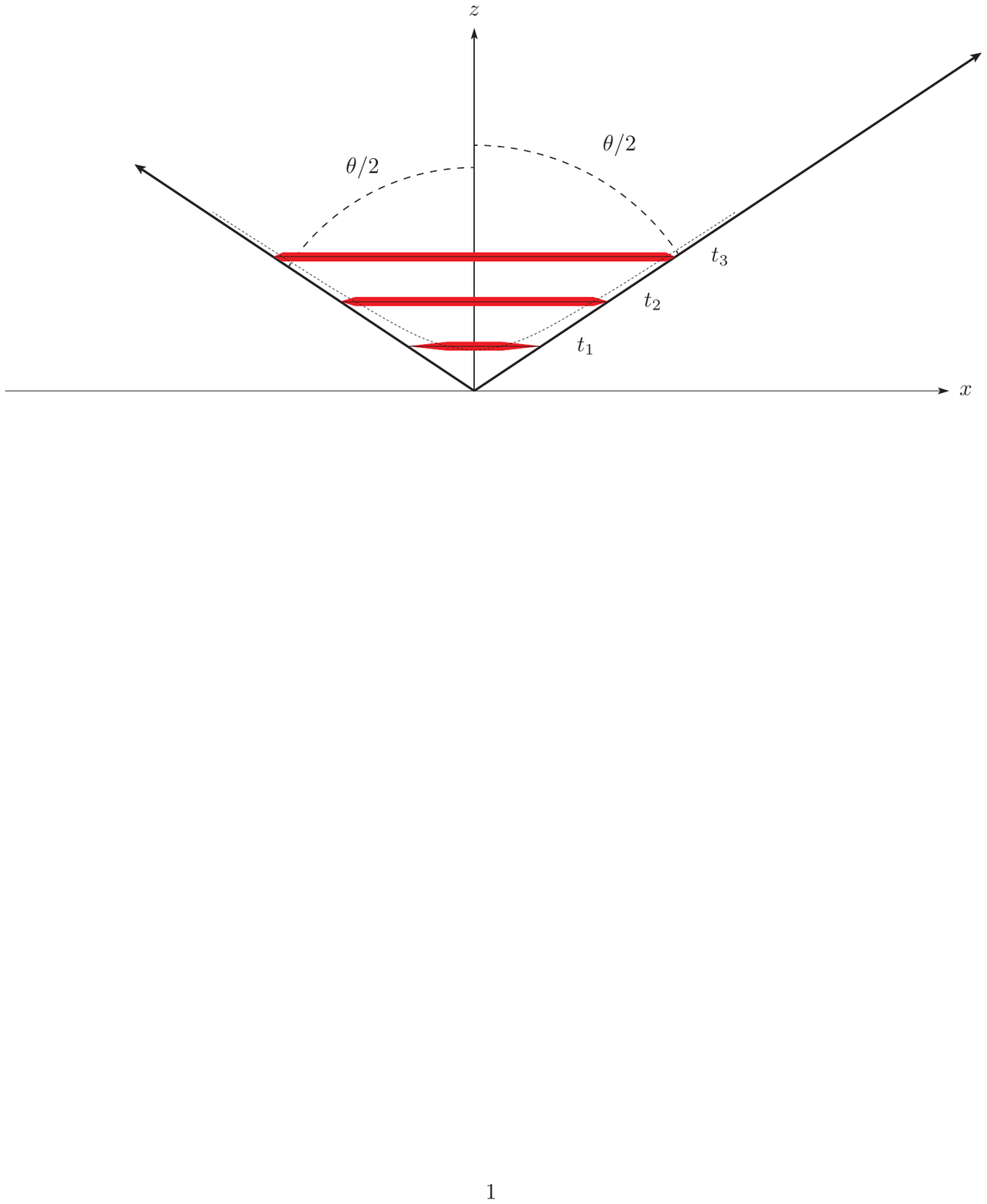}    
    \caption{\label{fig:onepiece}The time-evolution of a string piece
      between two partons in an arbitrary Lorentz frame. The system is
      rotated so that the momenta (the magnitude of which are
      indicated by the lengths of the arrows) of the of the partons
      are in the $x-z$ plane with equal but opposite angle w.r.t.\ the
      $z$-axis. The horizontal lines indicate the extension of the
      string horizontally and transversely at different times steps.
      The dotted curve indicates the hyperbola where in $x$ the proper
      time of the sting piece is equal to $R_S$ for a given
      $z=t\costh$. }
  }

If we instead go to an arbitrary Lorentz frame we can also obtain a
simple picture by rotating the partons so that they lie in the $x-z$
plane with the same but opposite angle $\theta/2$ with the $z$-axis,
as shown in \figref{fig:onepiece}. Here the string ends still move by
the speed of light and the position of string ends are given by
$x_\pm(t)=(t;\, \pm t\sinth,\, 0,\, t\costh)$. 
A straight relativistic string is boost invariant and has no longitudinal
momentum (similar to a homogeneous electric field). The energy and transverse
momentum are given by $\diff{E}/\diff{z} = \kappaS/\sqrt{1-v^2}$ and $\diff{p_\perp}/\diff{z} =
v_\perp \, \diff{E}/\diff{z}$. The string in \figref{fig:onepiece}  is therefore still a
straight line; it is just boosted transversely with velocity $v=\cos(\theta/2)$. 

\FIGURE[t]{
	\centering
	\begin{minipage}[c]{0.48\linewidth}
          \includegraphics[width=0.85\textwidth]{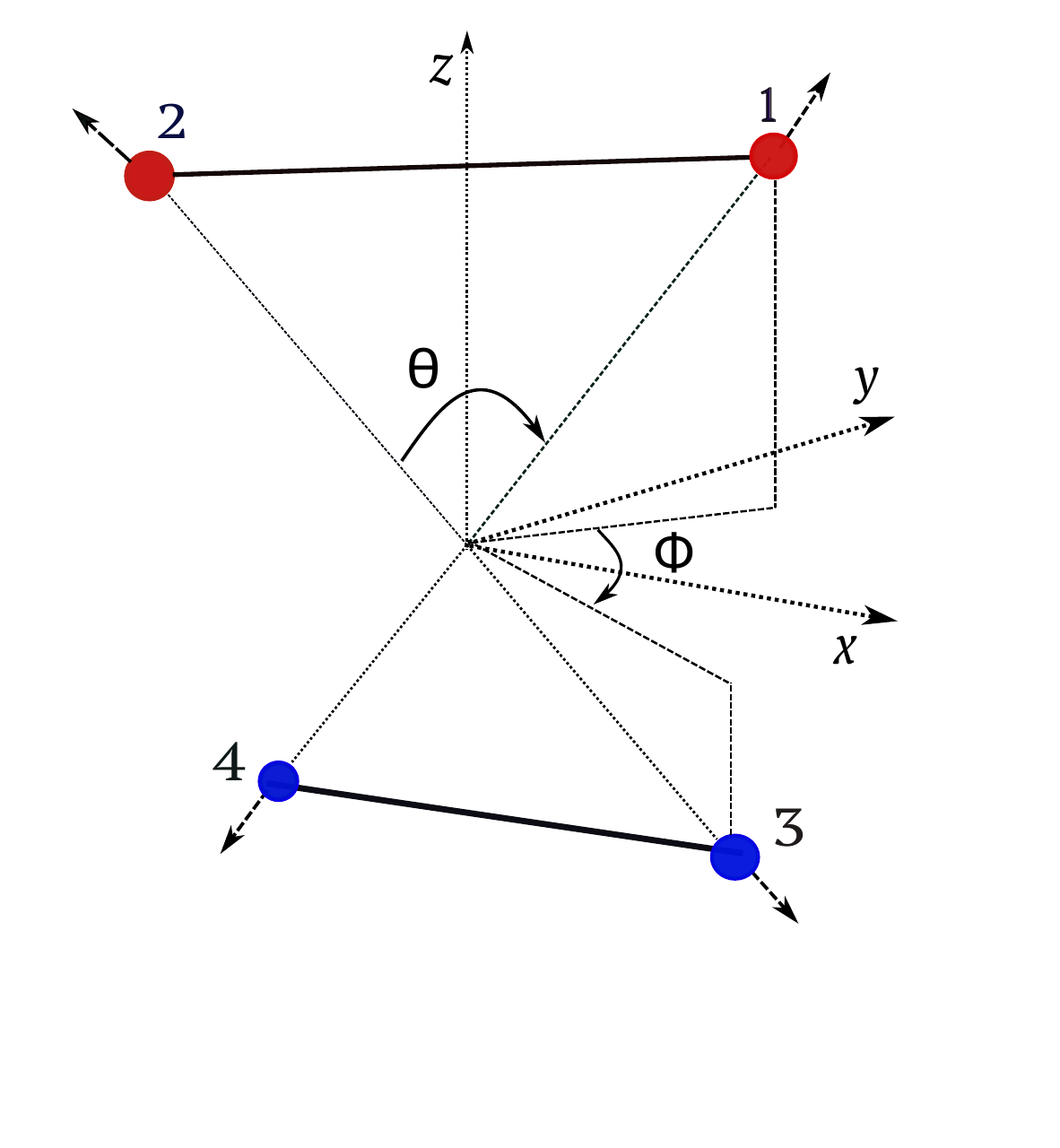}
	\end{minipage}%
        \begin{minipage}[c]{0.48\linewidth}
          \includegraphics[width=1.0\textwidth]{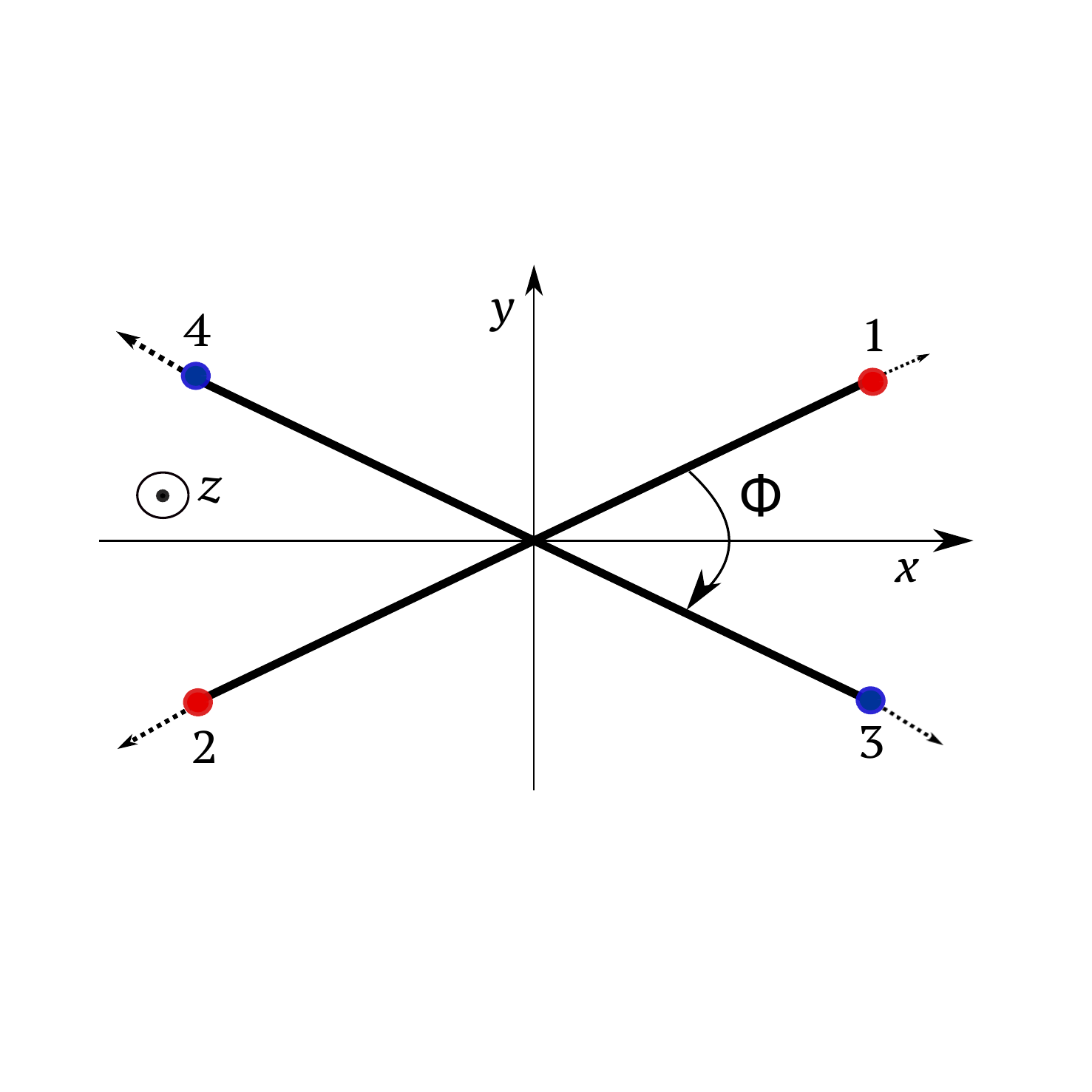}
	\end{minipage}
	\vspace*{-1.3cm}
	\caption{Schematic diagrams of two strings in their parallel
          frame.  On the left, $\theta$ is the opening angle between
          the partons constituting each string, while on the right we
          show the skewness angle, $\phi$ in the projection on the
          $x-y$-plane of the two string pieces. }
	\label{fig:parallelframe}
}

If we now want to study the interaction between two arbitrary string
pieces, it is not possible to find a Lorentz frame where both these
are always parallel to the $x$-axis, but it turns out that it is
possible to find a frame where both always lie in parallel planes,
perpendicular to the $z$-axis. To see this, we first we rotate the
string piece in \figref{fig:onepiece} an angle $\phi/2$ around the
$z$-axis. Then we want another string piece with angle $\pi-\theta/2$
w.r.t.\ the the $z$-axis but rotated an angle $-\phi/2$. We then have
the situation shown in \figref{fig:parallelframe} where the four
partons have momenta (using pseudorapidity instead of the polar
angle, to simplify the notation)
\begin{eqnarray}
\label{eq:pT}
p_1 &=& p_{\perp1} \left( \coshe; \phantom{-}\cosph, \phantom{-}\sinph, \phantom{-}\sinhe \right), \nonumber \\
p_2 &=& p_{\perp2} \left( \coshe; -\cosph,    -\sinph, \phantom{-}\sinhe\right), \nonumber  \\
p_3 &=& p_{\perp3} \left( \coshe; \phantom{-}\cosph,  -\sinph, -\sinhe \right), \nonumber  \\
p_4 &=& p_{\perp4} \left( \coshe; -\cosph,  \phantom{-}\sinph, -\sinhe \right).
\end{eqnarray}
Here we have six unknown quantities and, for four massless momenta we
can construct six independent invariant masses, $s_{ij}$. This means
that for any set of four massless particles we can (as long as no two
momenta are completely parallel) solve for the transverse momenta
\begin{equation}
p^2_{\perp1} = \frac{s_{12}}{4} \sqrt{\frac{s_{13}s_{14}}{s_{23}s_{24}}},\,\,\,
p^2_{\perp2} = \frac{s_{12}}{4} \sqrt{\frac{s_{23}s_{24}}{s_{13}s_{14}}},\,\,\,
p^2_{\perp3} = \frac{s_{34}}{4} \sqrt{\frac{s_{13}s_{23}}{s_{14}s_{24}}},\,\,\,
p^2_{\perp4} = \frac{s_{34}}{4} \sqrt{\frac{s_{14}s_{24}}{s_{13}s_{23}}},
\end{equation}
and the angles 

\begin{eqnarray}
\cosh \eta&=&\frac{s_{13}}{4 p_{\perp1}p_{\perp3}} + \frac{s_{14}}{4 p_{\perp1} p_{\perp4}}, \\
\cos \phi&=&\frac{s_{14}}{4 p_{\perp1} p_{\perp4}} - \frac{s_{13}}{4 p_{\perp1} p_{\perp3}} . 
\end{eqnarray}
We note that there is a mirror ambiguity in the solution, but apart
from that we can now construct a Lorentz transformation to take any
pair of string pieces to the desired frame, which we will call their
\textit{parallel frame}.

The sketches in \figref{fig:parallelframe} show the case where the
four partons are produced in the same space--time point, which in
general is not the case. The partons from the shower and MPI-machinery
in PYTHIA8 are all assigned a space--time positions $(0; bx , by , 0)$
in the lab frame assuming the standard picture that at t = 0 they are
packed together at z = 0 and only with transverse separation. When a
pair of string pieces are Lorentz transformed into their parallel
frame, we assume that their respective production points are a simple
average of the positions of the parton ends, giving us a
$p^a_0=(t^a_0;x^a_0,y^a_0,z^a_0)$ for the string piece moving along
the $z$-axis and the corresponding $p^b$ for the piece going in the
other direction.  From this we get for any given time, $t$, in the
parallel frame that the string piece travelling along the $z$-axis has
the length, $2(t-t^a_0)\sinth$, and lies in a plane transverse to the
$z$-axis. The string piece travelling in the opposite $z$ direction
will similarly have a length $2(t-t^b_0)\sinth$, which may be
different, but the string will still always lie in a plane
perpendicular to the $z$-axis. The endpoints of the two strings as a
function of time then become
\begin{eqnarray}
  \!\!p^{a\pm}(t)\!\!&=&\!\!\!\left(t;\,x^a_0 \pm (t-t_0^a)\sinth\cosph,\,y^a_0\pm (t-t_0^a)\sinth\sinph,
                 \, z^a_0  + (t-t_0^a)\costh\right)\nonumber\\
  \!\!p^{b\pm}(t)\!\!&=&\!\!\!\left(t;\,x^b_0 \pm (t-t_0^b)\sinth\cosph,\,y^b_0\mp (t-t_0^b)\sinth\sinph,
                 \,z^b_0 - (t-t_0^b)\costh\right)
\label{eq:stretch}
\end{eqnarray}
We see now that the distance between the two planes will change linearly with
time as $\Delta_z(t)=z_0^a-z_0^b -t_0^a+ t_0^b+ 2t\costh$.

Besides the string motion we are also interested in the radius of the
string. As indicated in \figref{fig:onepiece} this radius is not
constant along the string, but depends on the proper time of a point
on the string. As the partons are assumed massless, the endpoints of
the string always have zero proper time and the colour field there has
not had time to spread out transversely. Looking at a point on the
$x$-axis, $\bar{x}$, we can easily find the proper time
\begin{equation}
  \label{eq:eigentime}
  \tau(t,\bar{x})=\sqrt{(t-t_0)^2\sin^2\frac{\theta}{2}-(\bar{x}-x_0)^2-\left((\bar{x}-x_0)\tan\frac{\phi}{2}-y_0\right)^2}.
\end{equation}

The radius of the string will then vary linearly with $\tau$, from
zero in the ends until it reaches the final equilibrium radius,
$R_S=\tauS\lesssim1$~fm. After this the string's width is fixed (as
is indicated in \figref{fig:onepiece}) until it ultimately brakes,
which on the average happens at $\tauH\lesssim2$~fm. In our
implementation we have chosen to only allow the shoving to take place
between string pieces at points in the parallel frame where both
strings have proper times between $\tauS$ and $\tauH$.

Clearly there may also be shoving between string pieces where they
have not yet reached their maximum radius, $R_S$. From the derivation
of \eqref{eq:force} we see that the force will be larger and the range
will be smaller. In our current implementation it is possible to set
$\tauS<R_S$ to allow for shoving also in these region, but the
force is still given for $R=R_S$ in \eqref{eq:force}. This can therefore only
give an indication of the effect, and we have to postpone a quantified study
to a later publication.

The shoving naturally stops when the string breaks, but there is a
grey-zone after the string breaks and before the hadrons are fully
formed where one could imagine that the string pieces in the hadrons
being formed will still repel each other. Also, after the hadrons are
fully formed we expect final state interactions between them. In this
article we will not investigate final-state hadron interactions,
although it is able to produce a sizable flow signal in \PbPb~collisions 
with \angantyr initial conditions \cite{daSilva:2020cyn}, and has recently been 
implemented in \pytppp \cite{Sjostrand:2020gyg}. As indicated by the discussion
about the inherent ambiguity in defining a ``hadronization time'' in \sectref{sec:lundstring},
defining a transition between a string-dominated final state at early times, and
a hadron dominated final state at late times, will require scrutiny.

Another mechanism that reduces the shoving is when the endpoints have
limited momenta, $k_x$ in the $x$-direction in the parallel frame. A
parton loses momentum to the string at a rate $t\kappaS$, and a
gluon being connected to two string pieces loses momentum at twice
that rate. After a time $t=|k_x|/2\kappaS$ a gluon will therefore
have lost all its momentum to the string and turn back, gaining
momentum from the string in the opposite direction. As explained in
\sectref{sec:lundstring} (\figref{fig:threejet}), this means that a
new string region opens up which is then not in the same parallel
$x-y$ plane. In this article we do not treat this new string region,
but will comment on them in the following sections.

\subsection{Generating the shoving}
\label{sec:generating-shoving}

We now want to take the force between two string pieces in
\eqref{eq:force} and apply it in the parallel frame. In
\citeref{Bierlich:2017vhg} everything was done in the laboratory frame
and all strings were assumed to be parallel to the beam, and there
generation of the shove was done by discretizing time and rapidity,
calculating a tiny transverse momentum exchange in each such
point. Here we instead use the parallel frame and discretize the
transverse momentum into tiny fixed-size \textit{\pushes}.

Going back to the case where we have two completely parallel strings
we have from the expression for the force from \eqref{eq:force} as
\begin{equation}
\label{eq:the-force}
\frac{\diff{p_\perp}}{\diff{t}\diff{x}}=f(d_\perp(t))
\end{equation}
and we get the total transverse momentum push on the strings as
\begin{equation}
\label{eq:total-pt}
\Delta p_\perp =\int \diff{t} \int \diff{x} f(d_\perp(t)),
\end{equation}
where we note that the integration limits in $x$ are
time-dependent. Now we will instead \push several times with a fixed
(small) transverse momentum, $\delta p_\perp$ according to some
(unnormalized) probability distribution $P(t)$, which would give us
the total push
\begin{equation}
\label{eq:total-push}
\Delta p_\perp = \int \diff{t} P(t) \delta p_\perp,
\end{equation}
and for small enough $\delta p_\perp$ we can make the trivial
identification
\begin{equation}
\label{eq:prush-prob}
P(t) = \frac{1}{\delta p_\perp}\int \diff{x} f(d_\perp).
\end{equation}

In any string scenario we can now order the \pushes in time (in the
parallel frame), and we can ask the question which of the pairs of
string pieces will generate a transverse \push of size
$\delta p_\perp$ first. This gives us a situation that looks much like
the one in parton showers where the question is which parton radiates
first. And just as in a parton shower we will use the so-called
Sudakov-veto algorithm \cite{Sjostrand:2006za,Lonnblad:2012hz}. The
main observation here is that the probability of nothing to happen
before some time $t$ exponentiates, so that the (now normalized)
probability for the first thing to happen at time $t$ is given by
\begin{equation}
\label{eq:Sudakov}
P(t)e^{-\int_{t_{\min}}^tdt'P(t')}.
\end{equation}
In the Sudakov-veto algorithm, one can then generate the next thing to
happen in all pairs of strings individually, and then pick the pair
where something happens first.  In addition, since $P(t)$ may be a
complicated function, one may chose to generate according to an
simplified overestimate, $\hat{P}(t) \ge P(t)$, for which generating
according to \eqref{eq:Sudakov} is easier. In this way we get a trial
time, $t_t$, for the first thing to happen and then for that
particular time calculate the true value of $P(t_t)$, and accept the
chosen time with a probability $P(t_t)/\hat{P}(t_t)$. If we reject, we
then know we have \textit{under}estimated the probability of
nothing haven happened before $t_t$, which means we can generate the
next trial time from $t_{\min}=t_t$ (in \eqref{eq:Sudakov}).

In our case we will make the overestimate by treating the strings as
being completely parallel, only separated in $z$ and overestimating
the limits in the $x$-integration. We then generate a time and a point
along the $x$ axis, calculate the actual repulsion force there, to get
the true probability for accepting the generated time.

The whole procedure can be seen as discretizing in time with a
dynamically sized time step with larger time steps where the force is
small, and vise versa. This is a very efficient way of evolving in
time, and efficiency is important since we will calculate several
\pushes in each pair of string pieces in an event, and in the case of
\AA, there may be up to $\mathcal{O}(10^4)$ string pieces per event at
the LHC.

\FIGURE[t]{
	\centering
	\begin{minipage}{0.32\linewidth}
		\includegraphics[width=1.2\textwidth]{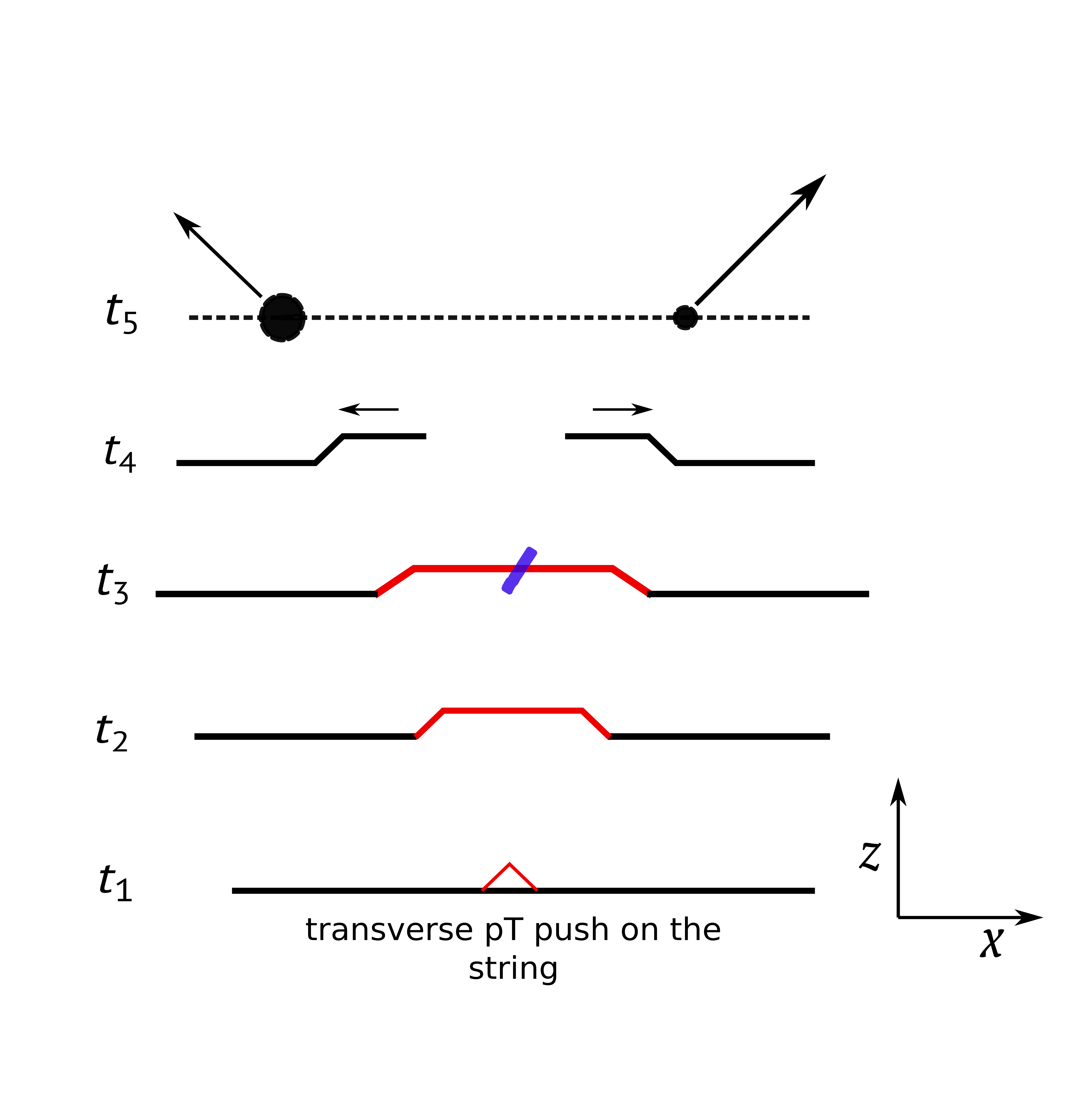}
	\end{minipage}\hfill\begin{minipage}{0.58\linewidth}
		\includegraphics[width=1.0\textwidth,bb=180 240 1750 1030]{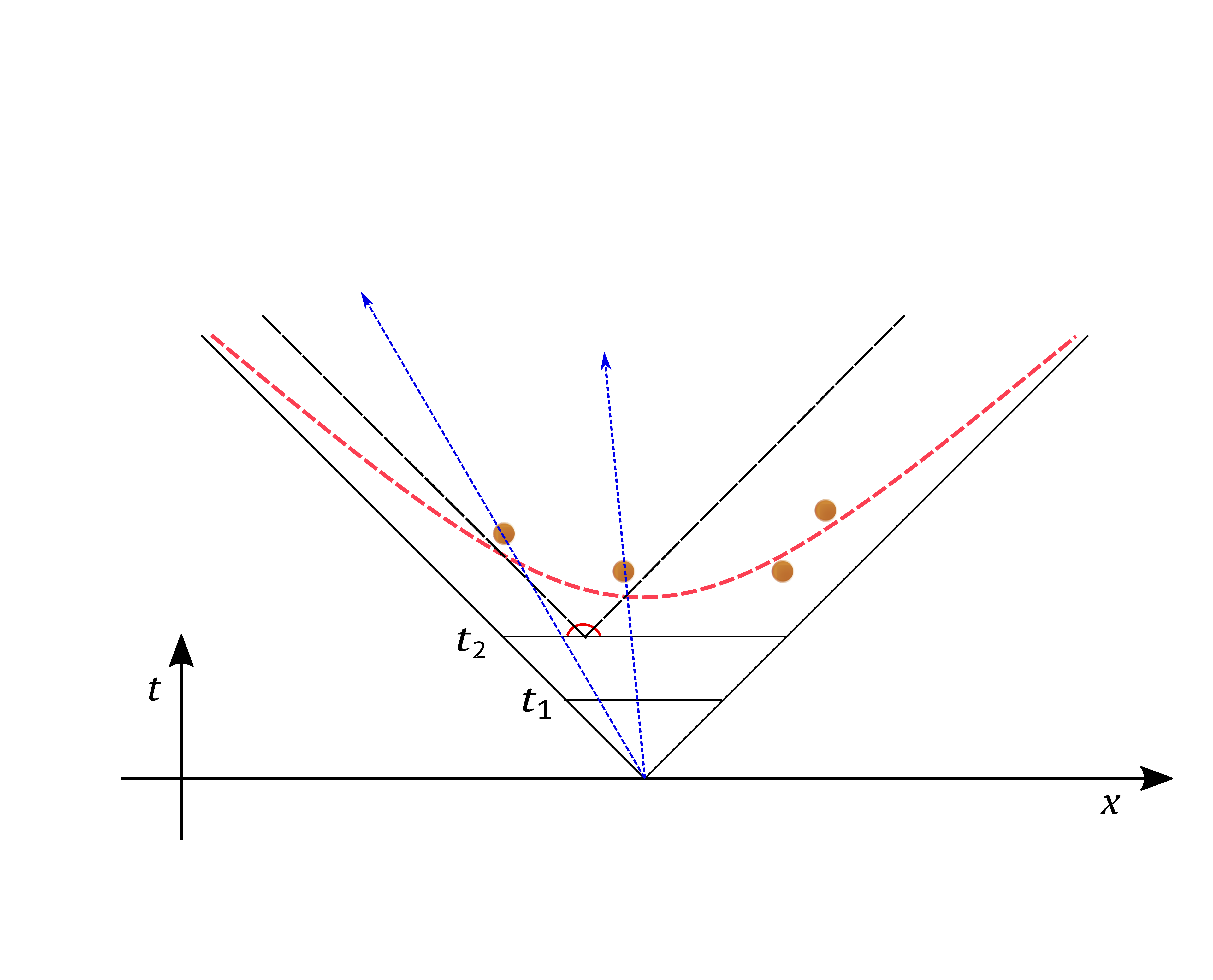}
	\end{minipage}
	
	\caption{The left figure illustrates how a \push deforms the geometry
		of a string piece from the time it is applied until the string
		breaks. The right figure illustrates how the final state hadrons
		that are will share the transverse momentum of a \push are chosen.}
	\label{fig:pushproagation}
}

\subsection{Transferring the \pushes to the hadons}
\label{sec:transf-push-hard}

Each time a \push has been included somewhere along a string piece,
the string is deformed slightly. It will correspond to a gluon kink on
the string, however, since the transverse momentum of this kink is
small, the $\delta p_\perp$ will soon start propagating along the
string as indicated in \figref{fig:pushproagation} (left). A hadron
that is produced along the string where there happens to be such a
kink will absorb the corresponding nudge in transverse momentum. In
practice this is done by finding the point where the proper time of
the kink is \tauH, the averaged hadronization time, and finding the
primary hadron with the corresponding pseudorapidity (which is
strongly correlated to the hyperbolic angle of the creation point, \cf\ \sectref{sec:density})
in the parallel frame, as sketched out in
\figref{fig:pushproagation} (right). In this way each \push\ is added
to two hadrons from each of the two interacting string pieces. It
should be noted that the fact that the two hadrons receive an extra
transverse momentum in the same direction, will also mean that they
come closer together in pseudorapidity. In addition, energy and
momentum conservation is achieved by the adjusting the longitudinal
momentum of the two hadrons separately in the two string pieces will
also (on average) pull the hadrons closer together in rapidity. In the
results below in \sectref{sec:aa-results} this becomes visible in the
overall multiplicity distribution.

It should be noted that the deformation of the string must be taken
into account, not only in the pair of string pieces where the \push
was generated, but in all pairs involving one of the string pieces,
triggering a recalculation of the next \push in all these pairs. To do
this in detail turns out to be forbiddingly time consuming, so instead
we estimate an average shift of a string piece only after a certain
number of \pushes (typically $\mathcal{O}(10)$) and distribute it
evenly along the string.
\footnote{The total number of \pushes is proportional to the square of
  number of string pieces, $N_S^2$. Requiring recalculation for all
  affected pairs after each \push increases the complexity to
  $\mathcal{O}(N_S^3)$. If we in addition would take into account the
  detailed geometry change for every previous \push, would make the
  complexity $\mathcal{O}(N_S^5)$, which would be forbiddingly
  inefficient.}

\section{Results for simple initial state geometries}
\label{sec:simple-aa}

The most widely employed models for describing the space--time evolution
of a final state interactions of heavy ion collisions as a QGP
(after the decay of a CGC as discussed in the introduction), are based on 
relativistic dissipative fluid dynamics (see \eg \citeref{Heinz:2013th} for a review). 
A key feature of such models, is that the observed momentum-space
anisotropy of the final state, originates from the azimuthal spatial anisotropy
of the initial state density profile. The final state anisotropy is quantified
in flow coefficients ($v_n$'s), which are coefficients of the Fourier expansion
of the single particle azimuthal particle yield, with respect to the event plane
($\Psi_n$) \cite{Voloshin:1994mz,Poskanzer:1998yz}:
\begin{equation}
	E\frac{\dthree{N}}{\dthree{p}} = \frac{1}{2\pi} \frac{\dtwo{N}}{p_\perp \diff{p_\perp} \diff{y}} 
	\left( 1 + 2\sum_{n=1}^\infty v_n \cos(n(\phi - \Psi_n)) \right).
\end{equation}
Here $E$ is the particle energy, $p_\perp$ the transverse momentum, $\phi$ the azimuthal
angle and $y$ the rapidity. In this section, we will explore the models' response to initial
state geometry in a toy setup without non-flow contributions from jets, \ie not real events, 
and thus we do not adapt the experimental calculation
procedures involving $Q$-vectors (see \sectref{sec:results} where we compare to real data),
but rather use the definition of the event plane angle from final state particles:
\begin{equation}
	\label{eq:event-plane}
	\Psi_n = \frac{1}{n} \arctan\left ( \frac{\langle p_\perp \sin(n \phi) \rangle}{ \langle p_\perp \cos(n \phi) \rangle}\right).
\end{equation}
Flow coefficients can then be calculated using only final state information as:
\begin{equation}
	\label{eq:vn}
	v_n = \langle \cos(n(\phi - \Psi_n)) \rangle.
\end{equation}

``Toy'' systems with known, 
simple input geometries are therefore better suited
for exploring the basic model dynamics. In \sectref{sec:pbtoy} we set up a toy model for high energy 
nuclear--nuclear collisions in which the shoving model can be applied, and in \sectref{sec:hydroexp} we
use the toy model to study the high-density behaviour of the shoving model. The parameters of the shoving
model are not tuned, but set at reasonable values of $g=0.5$, $R_S = \tauS = 1$ fm amd $\tauH = 2$ fm.

\FIGURE[t]{
  \centering
    \includegraphics[width=\textwidth]{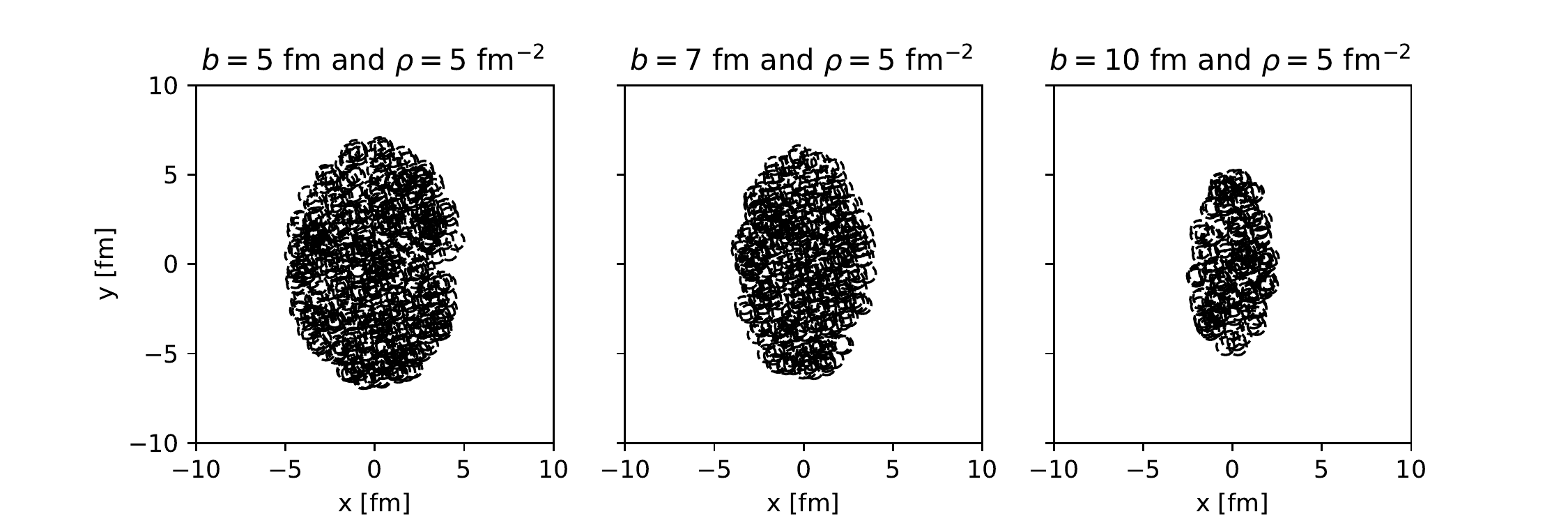}
  \caption{Examples of a sampled ellipse configuration in the Pb-Pb toy model, at three different impact parameters, with $\rho = 5$~fm$^{-2}$.}
  \label{fig:ellipse3}
}

\subsection{Isolating flow effects to $v_2$ in a toy model}
\label{sec:pbtoy}

In this and the following section, we restrict our study to systems with constrained straight strings. The strings are drawn 
between $\mrm{u}\bar{\mrm{u}}$ pairs, with cms energy of 15~GeV, a small Gaussian kick in 
$p_x$, $p_y$, and $p_z$ fixed by energy-momentum conservation.
This ensures that all strings will be stretched far enough in rapidity, that a study of final
state hadrons with $|\eta| < 1$ will not be perturbed by any edge effects. Final state hadrons 
studied in the figures, are all hadrons emerging from strings breakings (\ie no decays enabled)\footnote{We note 
that even a single string configuration can give rise to shoving effects, if the strings overlaps with itself.
Such configurations could possibly arise to the necessary degree in $e^+e^-$ collisions, though experimental
results so far have not shown any indication of flow \cite{Badea:2019vey,Abdesselam:2020snb}.}.

To study the model response in a heavy-ion like geometry, we set up a toy
geometry in the shape of an ellipse, drawn between two overlapping nuclei of $r_{\mrm{Pb}} = 7.1$~fm.
The ellipse has a minor axis ($\beta$) given by $2\beta = 2r_{\mrm{Pb}} - b$, and a 
major axis ($\alpha$) given by $2\alpha = \sqrt{4r^2_{\mrm{Pb}} - b^2}$, where $b$ is the impact parameter.
The elliptic overlap region is filled randomly with strings, given a certain density ($\rho$). Example events
for $\rho = 5\mrm{~fm}^{-2}$ are shown in \figref{fig:ellipse3}. We note that this
configuration is deliberately chosen to maximize $v_2$ (elliptical flow) at the expense of $v_3$ (triangular flow), 
though with a fluctuating initial state geometry, some $v_3$ will always be present
in order to study the response of the shoving model with minimal fluctuations.

\FIGURE[t]{
  \centering
  \begin{minipage}{1.0\linewidth}
    \begin{center}
    	\includegraphics[width=0.45\textwidth]{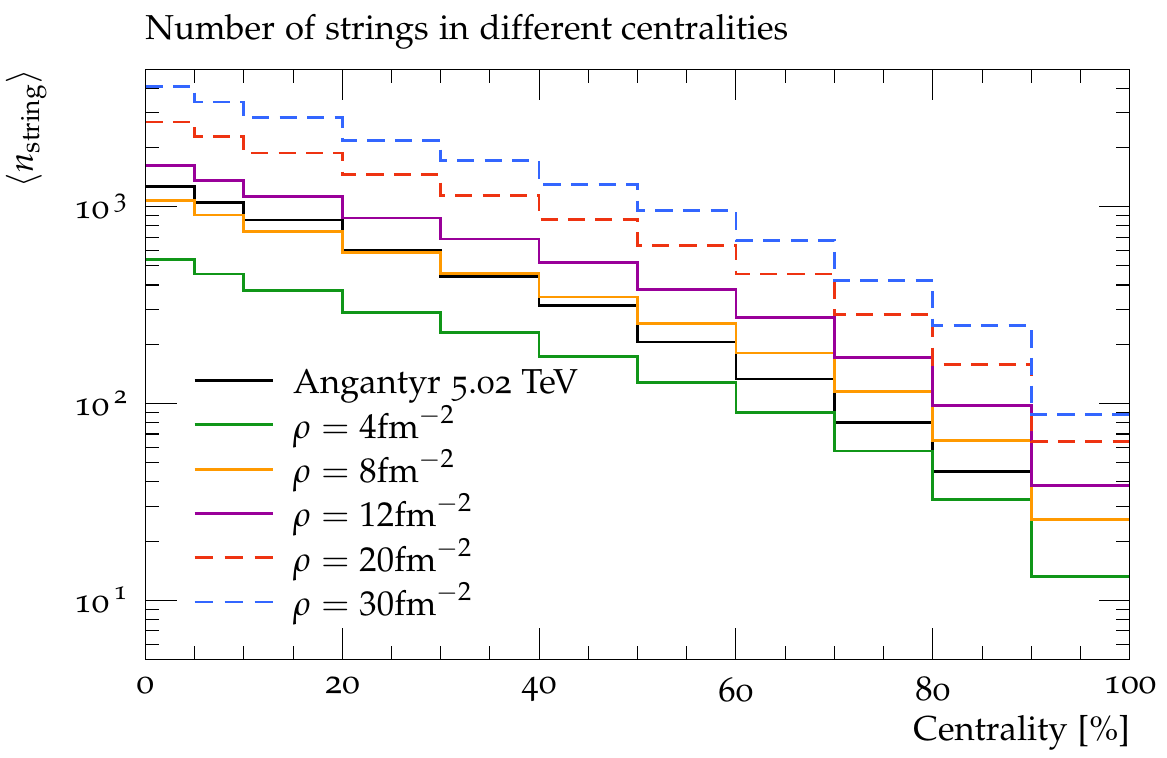}
    	\includegraphics[width=0.45\textwidth]{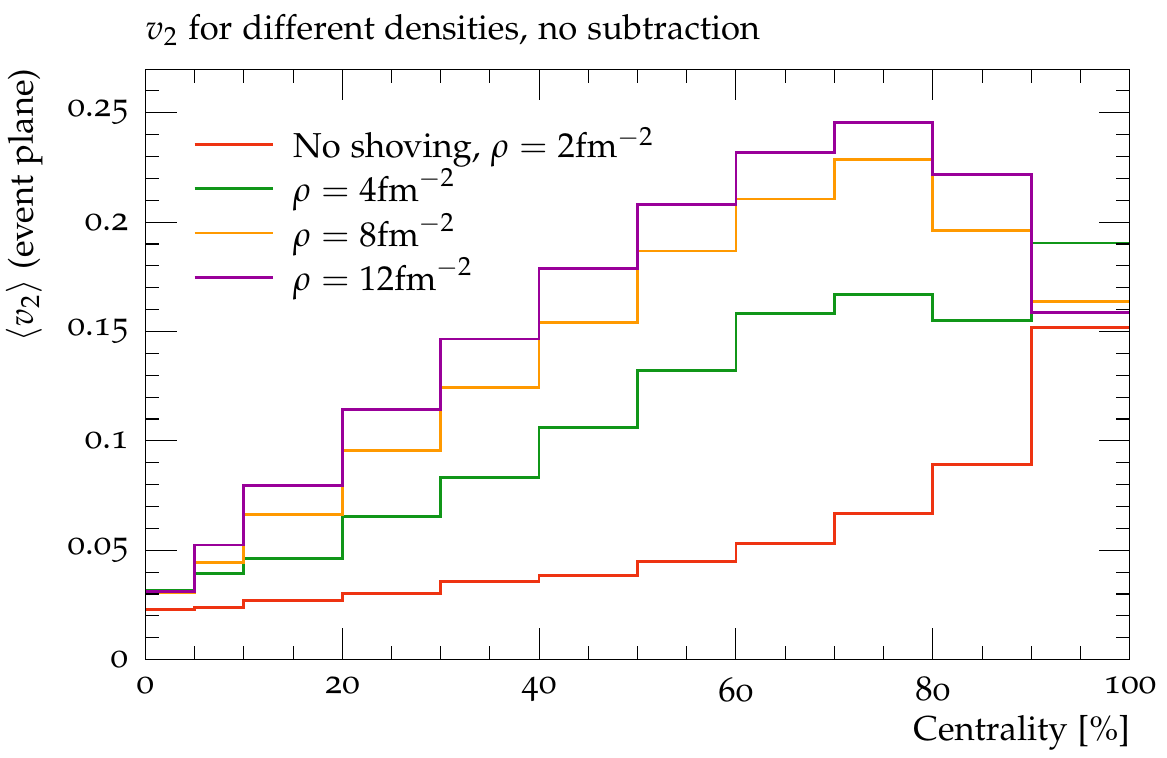}
    \end{center}
  \end{minipage}
  \caption{The number of string pieces (left) and $v_2$ (right) as function of centrality in the Pb-Pb toy model, for different values of $\rho$. In (right)
  compared to the number of strings generated by Angantyr at $\eta = 0$ for Pb-Pb collisions at $\sqrt{s_{\mathrm{NN}}} = 5.02$~TeV.}
  \label{fig:v2avg}
}
The initial anisotropy is given by the usual definition of the participant eccentricity \cite{Qiu:2011iv}, here employed on 
the strings:
\begin{equation}
	\label{eq:eccentricity}
	\epsilon_n = \frac{\sqrt{\langle r^2 \cos(n\phi) \rangle^2 + \langle r^2 \sin(n\phi) \rangle^2}}{\langle r^2 \rangle},
\end{equation}
where $r$ and $\phi$ are the usual polar coordinates of the string centers, but with the origin shifted to the
center of the distribution. It is possible to calculate higher order moments of \eqref{eq:eccentricity}, but here
we will be concerned only with $\epsilon^2_2\{2\} = \langle \epsilon^2_2 \rangle$, denoted by the short-hand $\epsilon_2$.
We begin by studying average quantities as a function of collision centrality, here defined by the impact parameter of the
two colliding nuclei, for exemplary

values of string density $\rho = \{4, 8, 12, 20, 30\}$~fm$^{-2}$.
In \figref{fig:v2avg} (left) we 
show the average number of strings in bins of centrality\footnote{The centrality is here defined by the impact parameter
between two colliding disks needed to give a certain elliptic geometry.}. The numbers are compared to the number of strings at $\eta = 0$ generated by the \angantyr model \cite{Bierlich:2018xfw} in Pb-Pb collisions at $\sqrt{s_{\mrm{NN}}} = 5.02$~TeV. The 
\angantyr model has been shown to give a good description of charged multiplicity at mid-rapidity in \AA~collisions \cite{Acharya:2018hhy},
and this comparison is therefore useful to provide a comparison to realistic string densities at current LHC energies.

\FIGURE[t]{
  \centering
  \begin{minipage}{1.0\linewidth}
    \begin{center}
    	\includegraphics[width=0.45\textwidth]{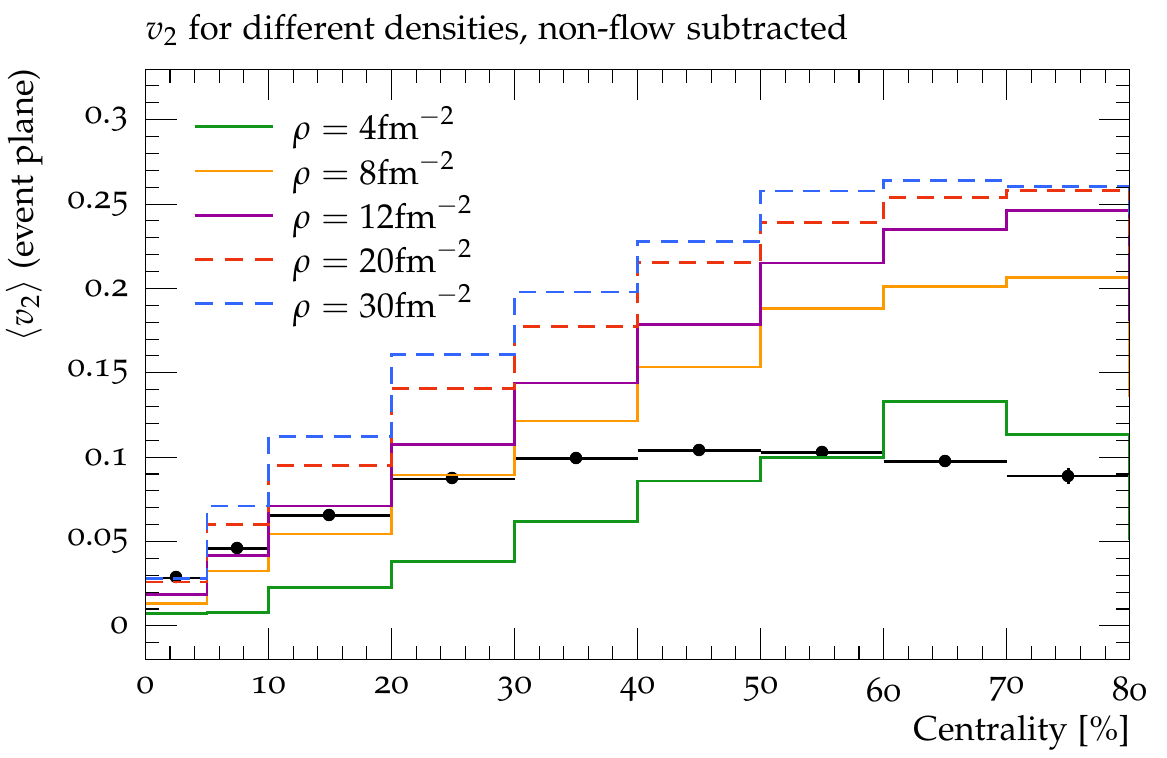}
    	\includegraphics[width=0.45\textwidth]{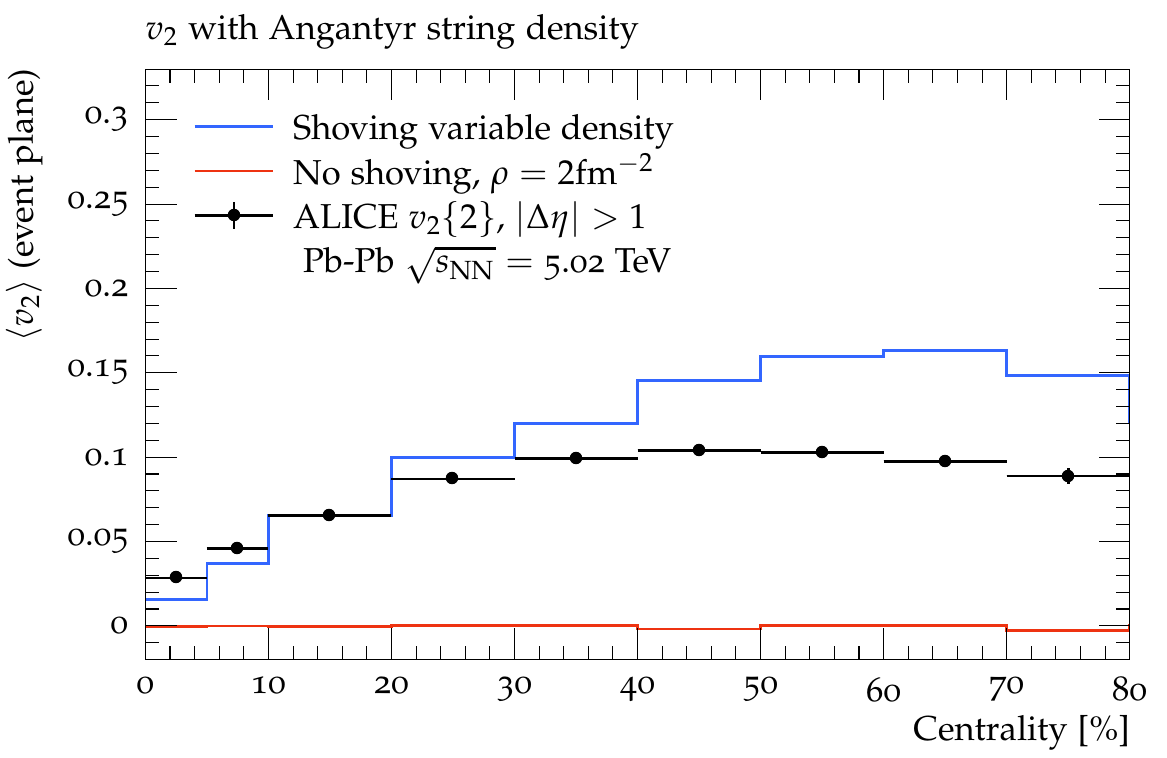}
    \end{center}
  \end{minipage}
  \caption{The $v_2$ flow coefficient with non-flow effects subtracted according to \eqref{eq:subtract} for different constant densities (left), and with string densities from \angantyr (right). Data from ALICE \cite{Adam:2016izf} added to guide the eye.}
  \label{fig:v2sub}
}

In \figref{fig:v2avg} (b), the average $v_2$ as a function of centrality is shown for the same densities as \figref{fig:v2avg} (a),
as well as for a reference without shoving, with $\rho=2$~fm$^{-2}$. 
Again, several observations can be made. First of all, for a fixed centrality $v_2$ will increase 
with increasing density. Secondly, the non-flow contribution is rather large -- in particular for peripheral centralities. This was also
noted in our recent paper on the \angantyr model \cite{Bierlich:2018xfw} using full Pb-Pb calculations. 
Third, for low string densities, the non-flow
contribution is so large, that it takes over in peripheral events, meaning that the curve exhibits no turn-over at a characteristic centrality.
Due to the toy nature of this setup, this cannot be taken as a firm prediction of the model, but should be investigated further in full simulations.

\FIGURE[t]{
  \centering
  \begin{minipage}{1.0\linewidth}
    \begin{center}
    	\includegraphics[width=0.45\textwidth]{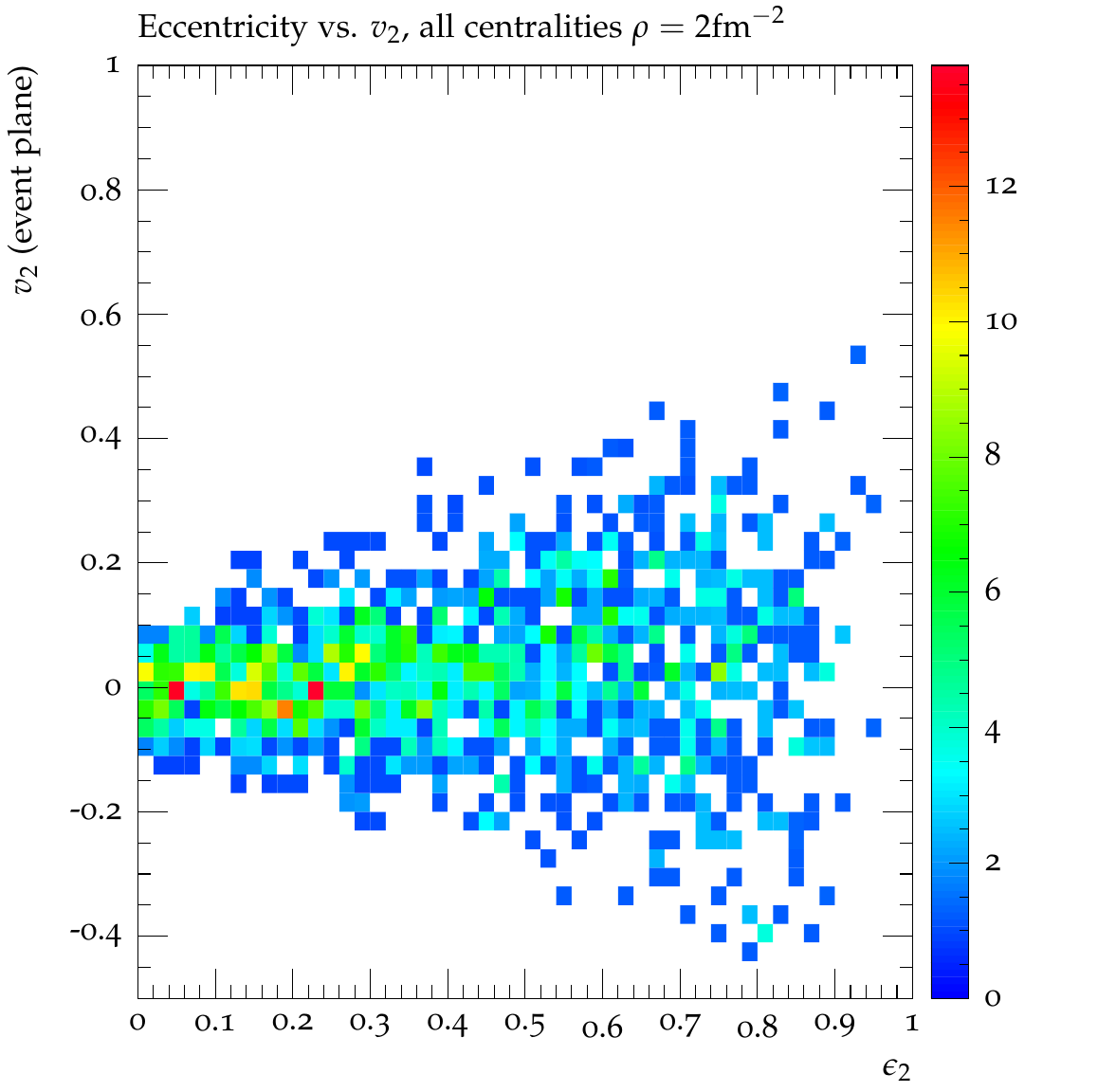}
    	\includegraphics[width=0.45\textwidth]{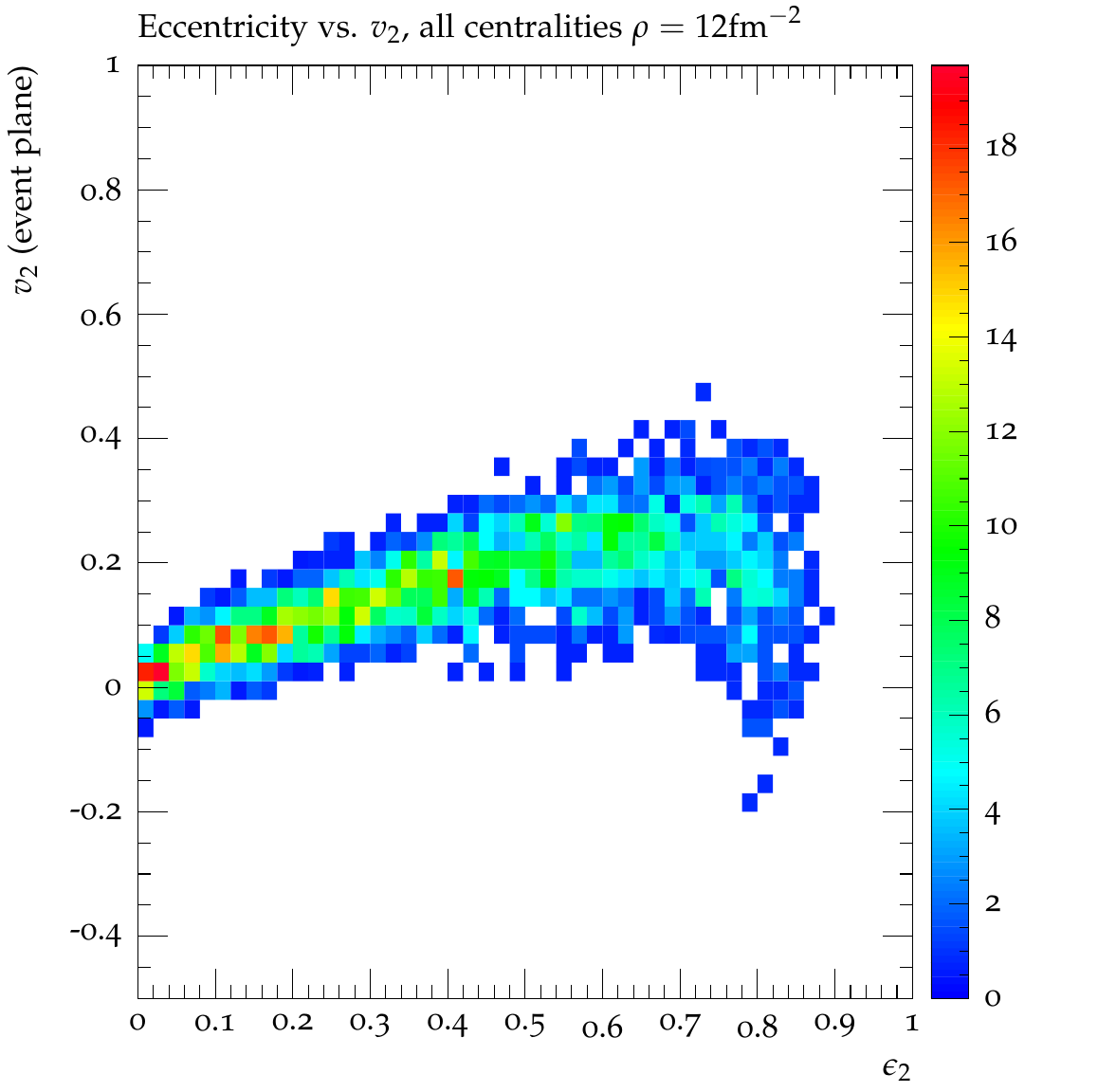}
    	\includegraphics[width=0.45\textwidth]{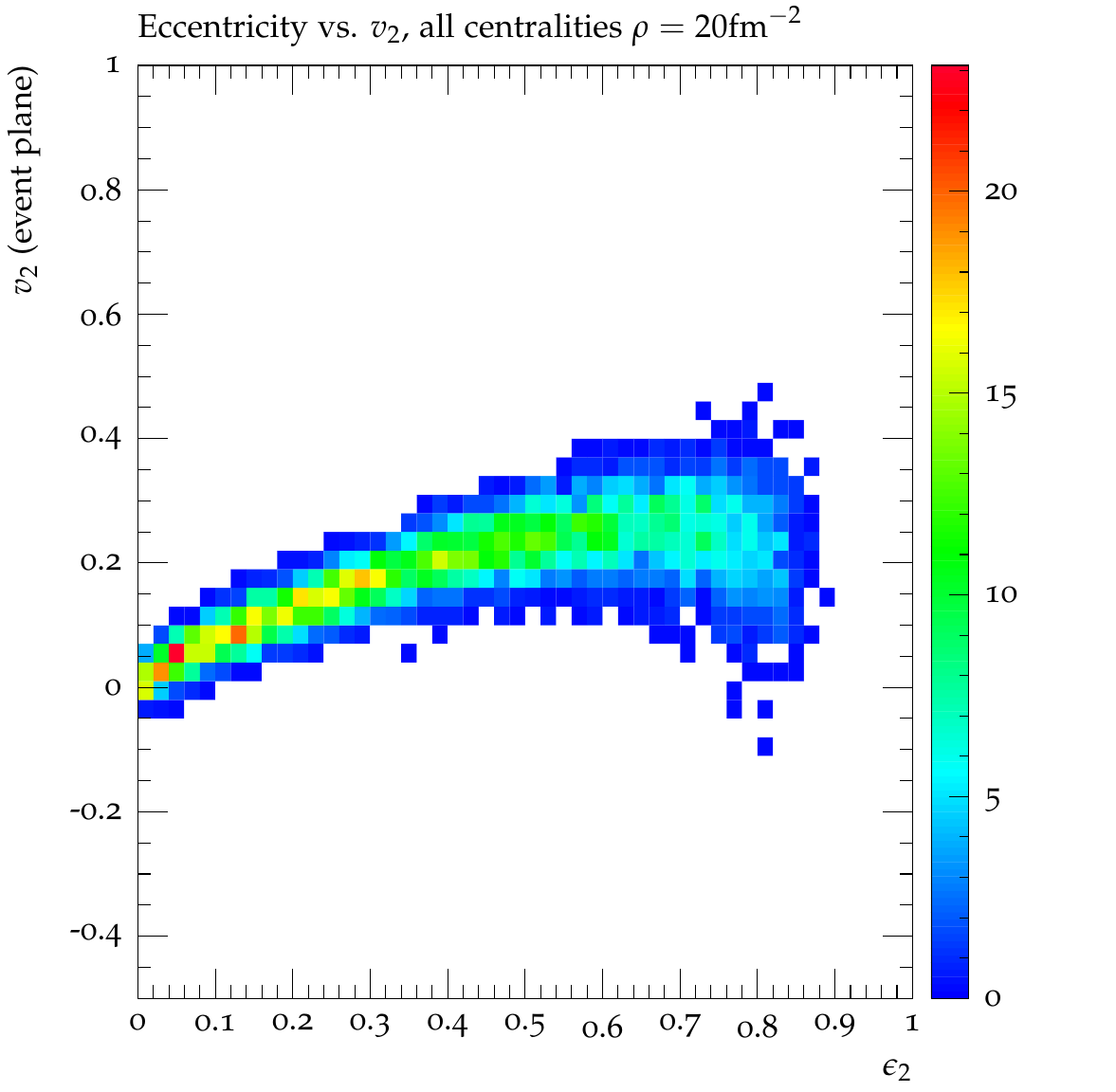}
    	\includegraphics[width=0.45\textwidth]{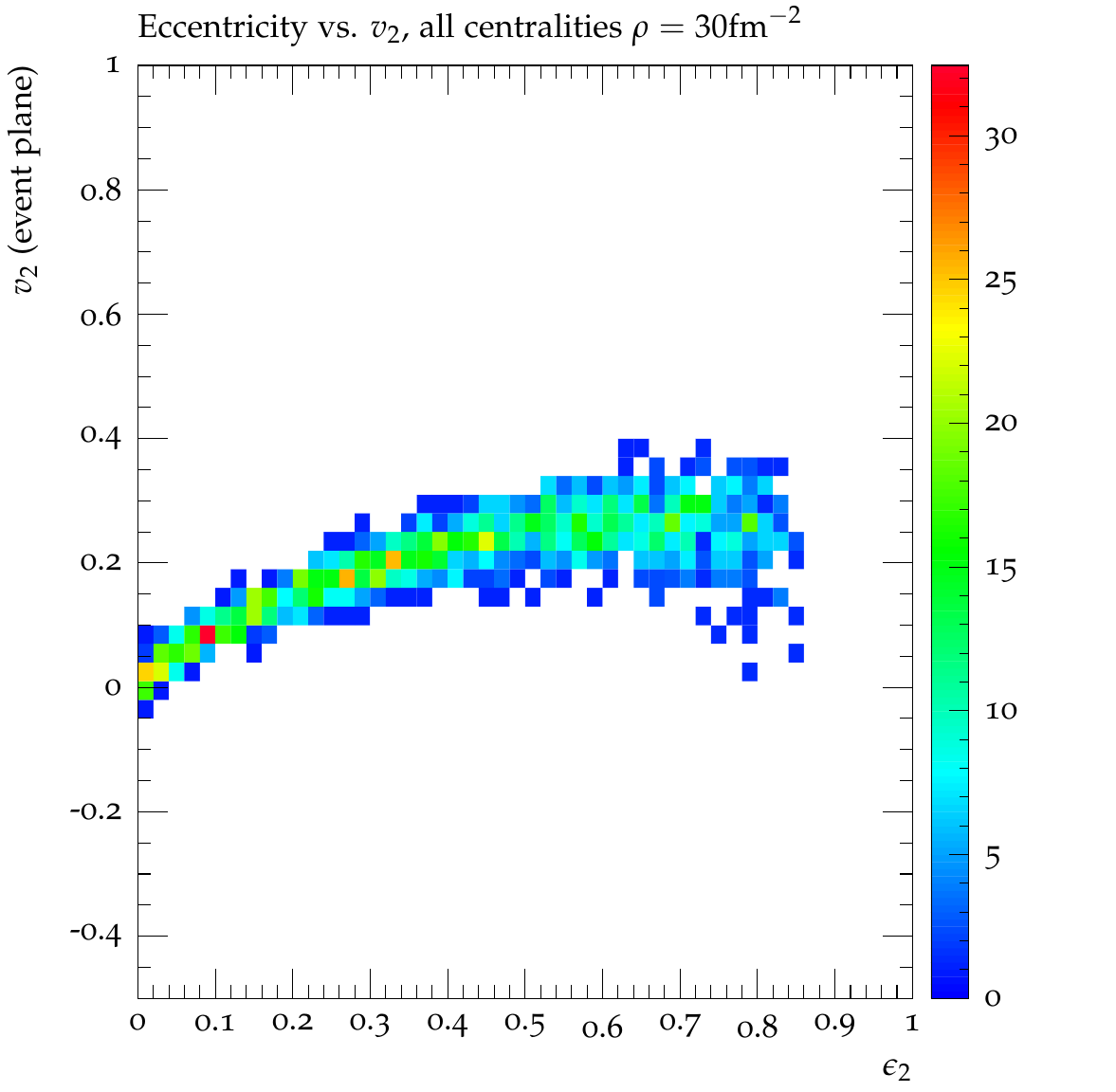}
    \end{center}
  \end{minipage}
  \caption{The correlation between $\epsilon_2$ and $v_2$ (non-flow subtracted) for four different densities, $\rho=2$~fm$^{-2}$ (upper left), $\rho=12$~fm$^{-2}$ (upper right),
  $\rho=20$~fm$^{-2}$ (lower left) and $\rho=30$~fm$^{-2}$ (lower right).}
  \label{fig:v2epsilon-all}
}

Since the goal of this section is to study the response to the elliptic geometry, the non-flow contribution will in 
the following be subtracted. To obtain this, each event is hadronized twice: once with shoving enabled, and once with shoving disabled.
For each of the two final states, $\Psi_n$ and $\cos(n(\phi - \Psi_n))$ is calculated, and the final, non-flow subtracted,
(inclusive) flow coefficients are given by:
\begin{equation}
	\label{eq:subtract}
	v_n^{\mrm{sub}} = \langle \cos(n(\phi - \Psi_n))_\mrm{shoving} - \cos(n(\phi - \Psi_n))_\mrm{no~shoving} \rangle_{\mrm{all~events}}.
\end{equation}

The flow coefficient $v_2$, with non-flow subtracted, is shown in \figref{fig:v2sub}.
In the left part of the figure, $v_2$ is shown for different constant densities. The higher density samples now
all have a characteristic turn-over at some centrality. It should be noted that the density intervals are not evenly spaced. Looking at $v_2$ in \figref{fig:v2sub}, the growth in the peak value decreases with higher density, like a saturation effect. Since final state multiplicity is proportional to the number of strings, the result in \figref{fig:v2sub} suggests that the shoving model dynamically predicts a steep rise in the peak value of $v_2$ at low densities (corresponding to low energies, with otherwise fixed geometry), which should then slow down.
The fact that string density is different at different centralities (as also shown in \figref{fig:v2avg} (a)), is used to construct \figref{fig:v2sub} (b). In this figure, string multiplicity for each toy-event are generated with \angantyr, and an elliptic initial condition of the \textit{same} impact parameter is constructed. This provides more realistic string densities at different centralities, and as it can be seen, this construction does a reasonable job at describing $v_2$ as a function of centrality. A no-shoving reference is also added to this figure, and as expected, it is consistent with zero. Finally it can be added, both $v_3$ and $v_4$ are, as could be expected from the engineered initial conditions, both compatible with zero after the subtraction, indicating that non-flow has been successfully removed. (Not shown in any figure here.)

\subsection{Towards the hydrodynamical limit at high string density}
\label{sec:hydroexp}

\FIGURE[t]{
  \centering
	    \includegraphics[width=1.0\textwidth]{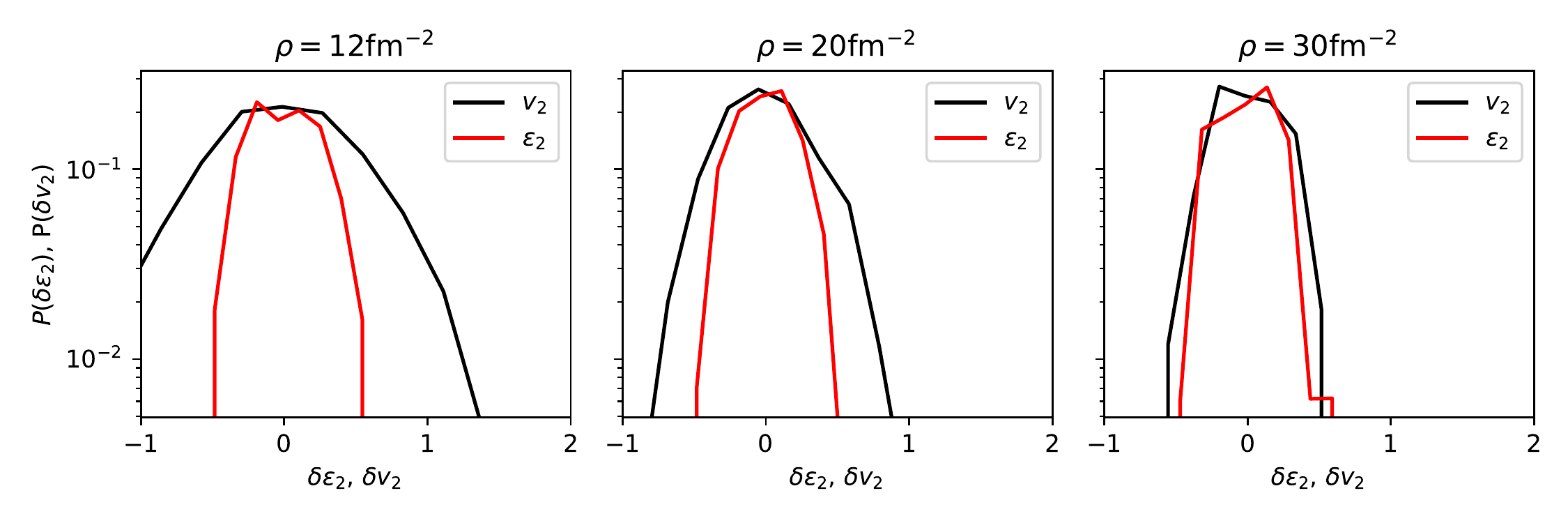}
  \caption{Probability distributions of $\epsilon_2$ and $v_2$ (non-flow subtracted) in the 20-30\% centrality bin, for three different
  values of $\rho$.}
  \label{fig:v2scaling}
}
As mentioned earlier, hydrodynamic calculations are often employed for heavy ion phenomenology, when it concerns interactions of the
final state, leading to collective flow. Whether or not the shoving model will, in the end, describe heavy ion data to a satisfactory 
degree, it is interesting to ask to what kind of hydrodynamics the shoving model will become in the thermodynamic limit. While analytic 
exploration of this question will be left to future studies, it is possible at this point to study similarities in the phenomenology
of the two approaches. An obvious route is to follow the paper by Niemi \etal \cite{Niemi:2012aj}, which studied the correlation between
flow coefficients and eccentricities, and noted that $v_2$ and $v_3$ have a strong linear correlation with $\epsilon_2$ and $\epsilon_3$.
In \figref{fig:v2epsilon-all}, we show the correlation between $\epsilon_2$ (from \eqref{eq:eccentricity}) and $v_2$, for four values of 
$\rho$. No centrality selection is performed, as the toy geometry of the system impose a strict relationship between eccentricity and
centrality defined by impact parameter, and binning in centrality would thus not add any information.
In the very dilute limit, $\rho = 2$~fm$^{-2}$, in \figref{fig:v2epsilon-all} (upper left),
it is seen that even though
$\langle v_2 \rangle$ in \figref{fig:v2sub} is non-zero, there is no strong correlation with $\epsilon_2$, and the distribution of $v_2$ is
rather wide. At intermediate densities $\rho = 12$ and $20$~fm$^{-2}$, in \figref{fig:v2epsilon-all} (upper right) and (lower left), the strong correlation 
appears, though not as narrow as in \citeref{Niemi:2012aj}. Finally for the densest considered state, $\rho = 30$~fm$^{-2}$ in 
\figref{fig:v2epsilon-all} (lower right) the correlation is, on average, similar to the more dilute ones, but more narrow.

To further study the scaling of $v_2$ with $\epsilon_2$, we study the scaled event-by-event variables:
\begin{equation}
	\delta \epsilon_2 = \frac{\epsilon_2 - \langle \epsilon_2 \rangle}{\langle \epsilon_2 \rangle} \mrm{,~~~and~~~} 
	\delta v_2 = \frac{v_2 - \langle v_2 \rangle}{\langle v_2 \rangle}.
\end{equation}
In \figref{fig:v2scaling} the distributions of $\delta \epsilon_2$ and $\delta v_2$, are shown for the densities 
$\rho = \{12, 20, 30\}$~fm$^{-2}$, in the 20-30\% centrality bin. It should be noted that, due to the purely elliptical
sampling region, the shape of the distribution cannot be compared directly to those of \citeref{Niemi:2012aj}. The
main conclusion is, however, clear. In the dense limit, in the rightmost panel, the two distributions are almost
identical. This means that the shoving model, in this limit, reproduces a key global feature of hydrodynamics, namely
full scaling of final state (purely momentum space) quantities with the global initial state geometry.

A notable discussion pertains to the issue whether or not heavy ion collisions at RHIC and LHC energies, reach a 
high enough string density for the shoving model to behave like hydrodynamics, as indicated above. In \figref{fig:v2avg} (left),
it was shown that Pb-Pb collisions at LHC energies produce an amount of strings corresponding to toy model densities of around
10~fm$^{-2}$ for very central events, 8~fm$^{-2}$ for mid-central events (corresponding to the two leftmost panels of \figref{fig:v2scaling})
and less than 4~fm$^{-2}$ for the most peripheral events. Even though the toy model predicts scaling of $\langle v_2 \rangle$ with 
$\langle \epsilon_2 \rangle$, the fluctuations do not exhibit the same scaling. A further direct study of flow fluctuations in non-central
heavy ion collisions would thus be of interest both on the phenomenological and experimental side.

\section{Results with \angantyr initial states}
\label{sec:results}

In the previous section, we have shown that given a simple initial state of long, straight strings without any soft gluons,
the shoving mechanism can produce a response which (a) scales with initial state geometry in the same way as a 
hydrodynamic response, and (b) can produce flow coefficients in momentum space at the same level as measured in experiments.
In this section we go a step further, and present the response of the model given a more realistic initial string configuration,
as produced by the \angantyr model, which can be compared to data. In section \sectref{sec:thepicture} we described
several of the challenges faced when interfacing the shoving model to an initial state containing many soft gluons, which in particular
is the case in \AA collisions. Throughout this section we use the same canonical values of shoving model parameters as in the previous section. 

\subsection{Results in \pp collisions}
\label{sec:pp-results}

Already the original implementation of the shoving model was shown in \citeref{Bierlich:2017vhg} to give a satisfactory description of the \pp ``ridge''.
We will in this section focus on 
flow observables as calculated at high energy heavy ion experiments, \ie flow coefficients calculated using the generic framework formalism \cite{Bilandzic:2010jr,Bilandzic:2013kga},
in the implementation in the \rivet program \cite{Bierlich:2020wms}.

It is in principle possible to generate \pp events using the normal \pythia MPI model \cite{Sjostrand:1987su,Sjostrand:2004pf,Corke:2010yf}.
It would, however, be computationally
inefficient, since emphasis should be given to high multiplicity results. The results presented here therefore use the modifications
of the MPI framework presented as part of the \angantyr heavy ion model \cite{Bierlich:2018xfw},
notably the ability to bias the impact parameter selection towards very central
pp collisions,
and re-weighting back to the normal distribution. 

We first note, that while the shoving mechanism does not change the total multiplicity of an event, it will change the multiplicity in fiducial 
region measured by an experiment, because it will push particles from the unmeasured low-$p_\perp$ region to (measured) higher $p_\perp$.
It is therefore necessary to slightly re-tune the model parameters, to obtain a correct description of basic observables such as $p_\perp$ distributions
and total multiplicity.
In practise, the parameter $p_{\perp,0}$, which regulates\footnote{The divergence of the partonic $2\rightarrow 2$ cross section is regularized for $p_\perp \rightarrow 0$ by a factor $p_\perp^4 / (p_{\perp 0}^2 + p_\perp^2)^2$, and by using an $\alpha_s(p_{\perp 0}^2 + p_{\perp}^2)$. Tuning is done by \texttt{MultipartonInteractions:pT0Ref} which is the $p_{T 0}$ value for the reference CM energy (where \texttt{pT0Ref} = \texttt{pT0(ecmRef))}.} the $2\rightarrow 2$ parton cross section is increased from the default
value of 2.28 to 2.4. In \figref{fig:tuning} we show a comparison between \pythia (with the above mentioned impact parameter sampling) and \pythia + shoving, for a few standard minimum-bias observables,
compared
the charged $p_\perp$ distribution (left), the distribution of number of charged particles
(middle) and $\langle p_\perp \rangle (N_{ch})$ (right), all at
$\sqrt{s} = 13$~TeV. Data by ATLAS \cite{Aad:2016mok}, the analysis implemented in the \rivet framework \cite{Bierlich:2019rhm}.

\FIGURE[t]{
  \centering
  \begin{minipage}{1.0\linewidth}
    \begin{center}
	    \includegraphics[width=0.3\textwidth]{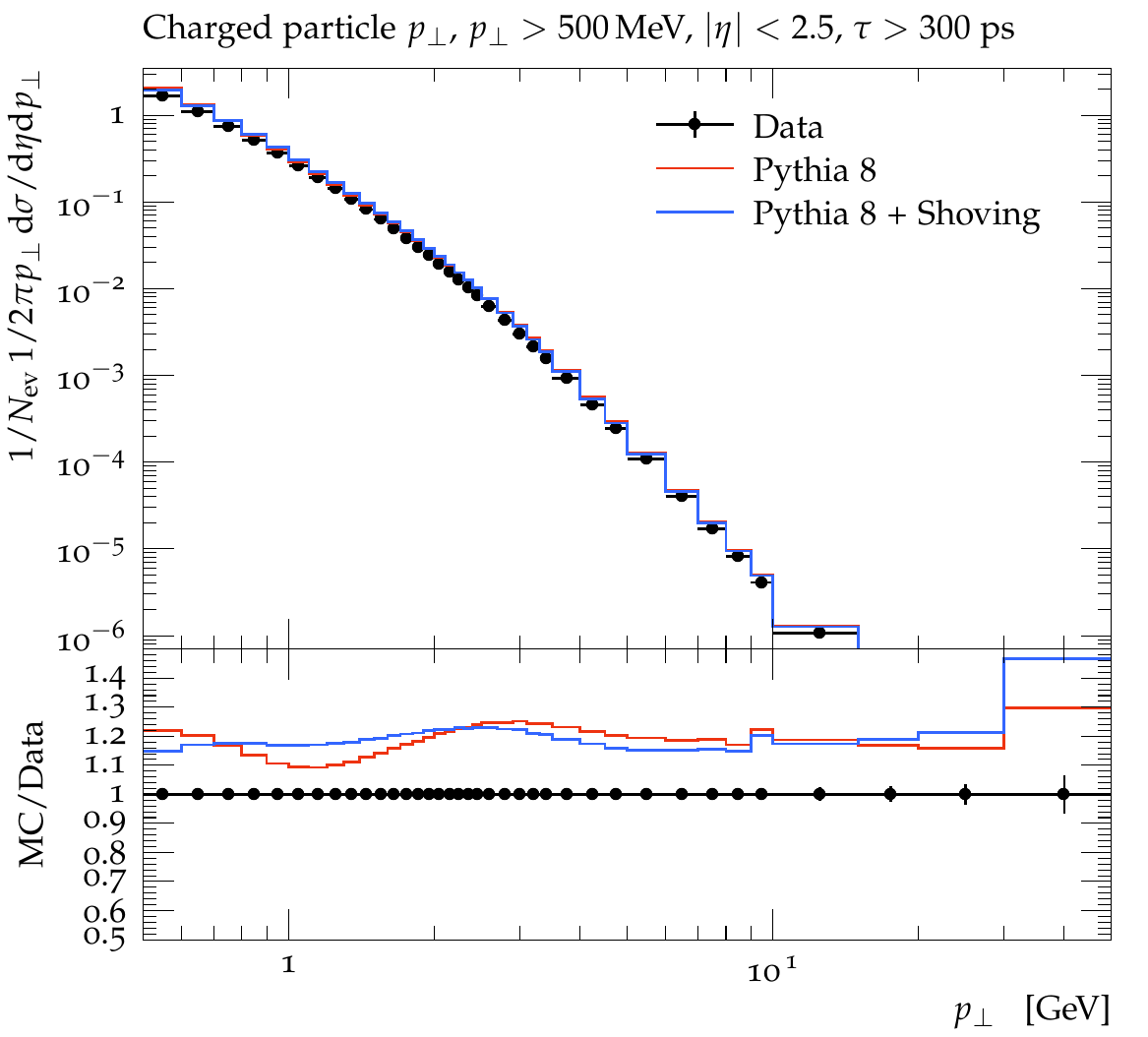}
	    \includegraphics[width=0.3\textwidth]{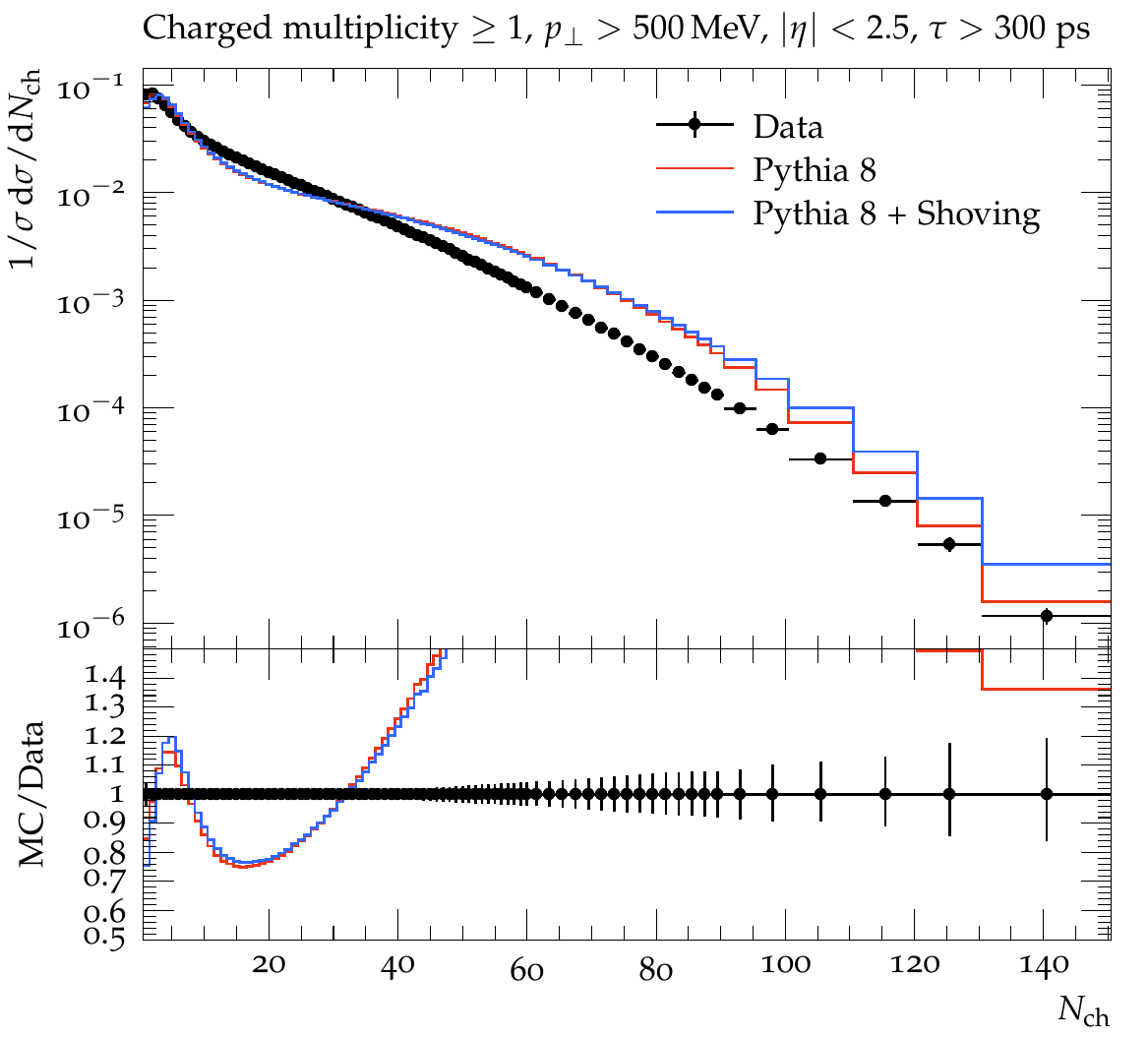}
	    \includegraphics[width=0.3\textwidth]{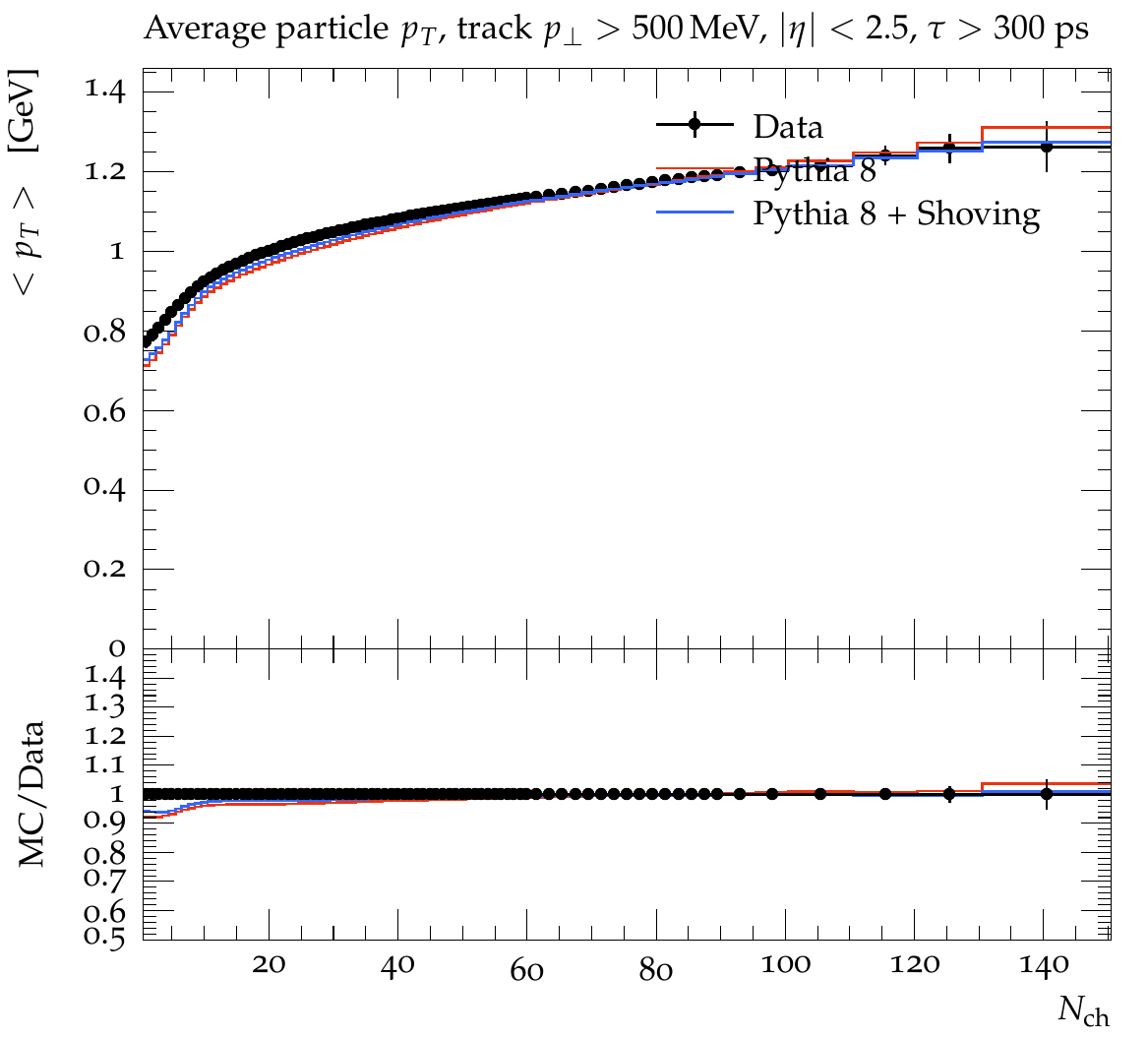}
    \end{center}
  \end{minipage}

  \caption{Comparison of \pythia (non-diffractive) to \pythia + shoving to basic $p_\perp$ (left), multiplicity (middle) and $\langle p_\perp \rangle$ distributions
  in \pp collisions at $\sqrt{s} = 13$~TeV to data by ATLAS.}
  \label{fig:tuning}
}

The agreement between simulation and data for the multiplicity distribution, is not as good
as normally expected from \pythia. This is expected, as only non-diffractive collisions were simulated. More interesting is the slight difference between the two simulations, where it
is clearly visible that in spite of the re-tuning, shoving still produce more particles in the high-$N_{ch}$ limit. In the $p_\perp$ distribution, it is seen that shoving has the effect
of increasing the spectrum around around $p_\perp = 1$~GeV. While the normalization if off (due to the exclusion of diffractive events in the simulation), shoving brings the shape of 
the low-$p_\perp$ part of the spectrum closer to data. Finally the $\langle p_\perp \rangle (N_{ch})$ is almost unchanged.

We now turn our attention to flow coefficients, and show $v_2$ calculated by two-particle correlations in \figref{fig:v2-pp}. In \figref{fig:v2-pp} (left) the multiplicity dependence of $v_2\{2\}$ with $|\Delta \eta| > 1.4$ is shown, and in \figref{fig:v2-pp} (right) the $p_\perp$-dependence of $v_2\{2\}$ ($|\Delta \eta| > 2$) in high multiplicity events is shown. Several conclusions can be drawn from the two figures.

\FIGURE[t]{
  \centering
  \begin{minipage}{1.0\linewidth}
    \begin{center}
	    \includegraphics[width=0.45\textwidth]{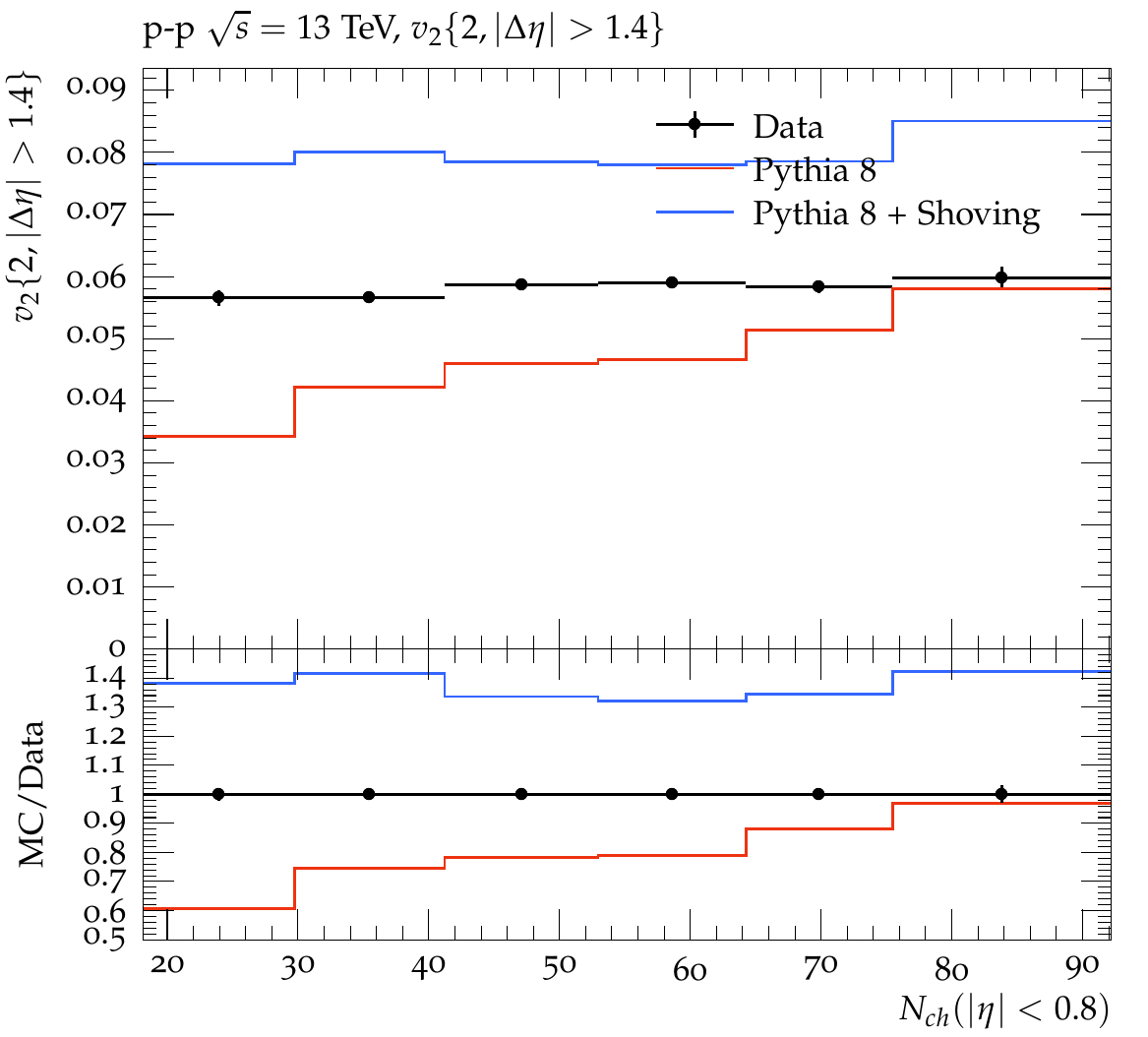}
	    \includegraphics[width=0.45\textwidth]{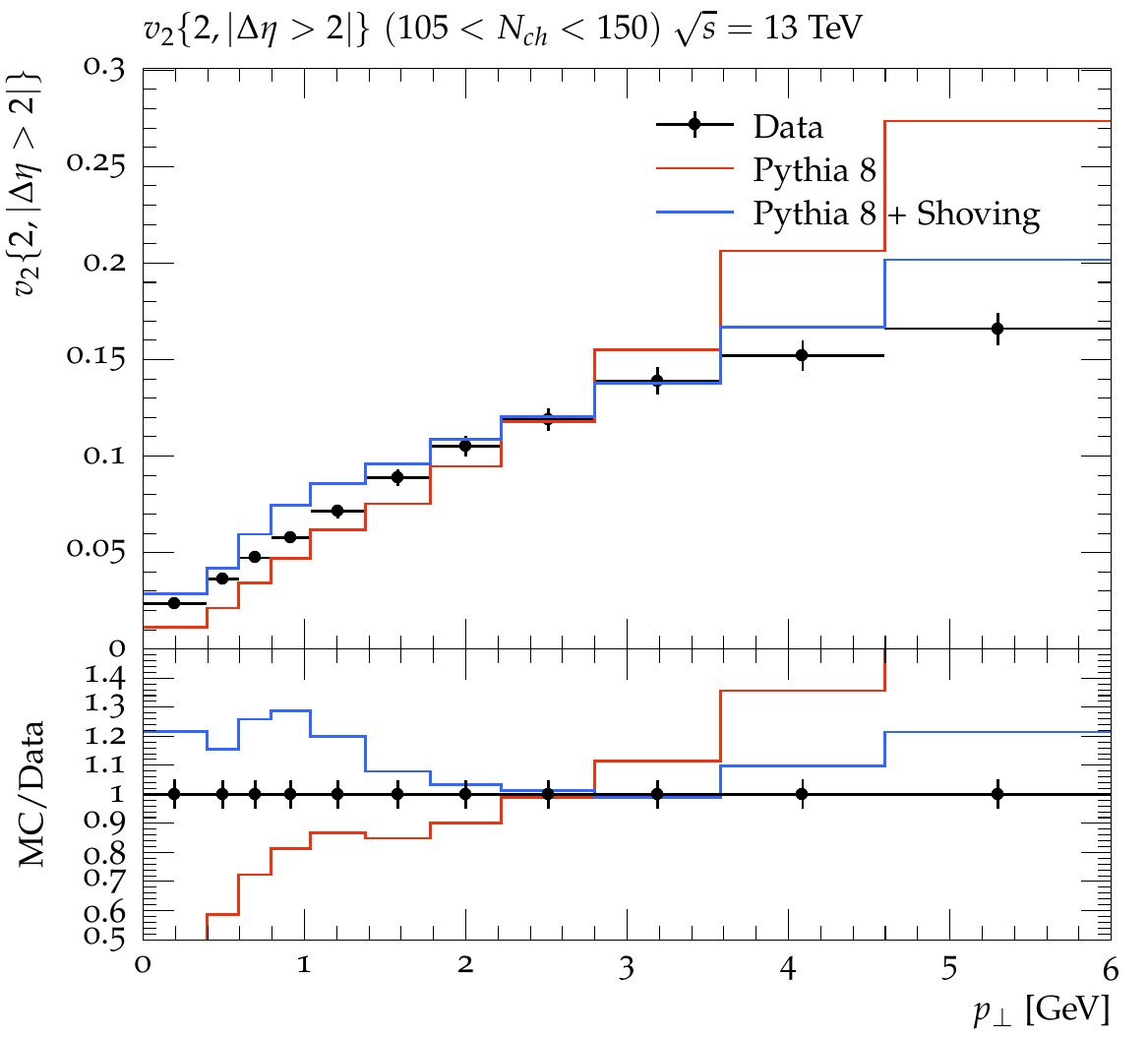}
    \end{center}
  \end{minipage}

  \caption{Comparison to $v_2\{2\}$ as function of multiplicity with ALICE high multiplicity trigger (left), and versus $p_\perp$ in high multiplicity events (right). Data from \pp collisions at $\sqrt{s} = 13$~TeV by ALICE \cite{Acharya:2019vdf}
      and CMS \cite{Khachatryan:2016txc}.}
  \label{fig:v2-pp}
}
First of all, it is seen that $v_2$ as a function of multiplicity (in \figref{fig:v2-pp} (left)) is too high with shoving enabled. We emphasize that the model parameters have not been tuned to reproduce this data, and
in particular that successful description of this data will also require a good model for the spatial distribution of strings in a \pp collision -- a point we will return to in a moment. We do, however, note that the additional $v_2$ added by the shoving model, persists
even with an $\eta$ separation of correlated particles (as $|\Delta \eta|$ cuts are applied),
a feature which separates the shoving model from \eg colour reconnection approaches, which have been pointed out to produce flow-like effects in \pp collisions \cite{Ortiz:2013yxa,Bierlich:2015rha,Bierlich:2018lbp}.
We also note that the $p_\perp$-dependence of $v_2$ is drastically improved,
as seen in \figref{fig:v2-pp} (right). In particular the high-$p_\perp$ behaviour of this quantity is interesting, as it decreases wrt.\ the baseline when shoving is enabled. 

\FIGURE[t]{
  \centering
  \begin{minipage}{1.0\linewidth}
    \begin{center}
	    \includegraphics[width=0.45\textwidth]{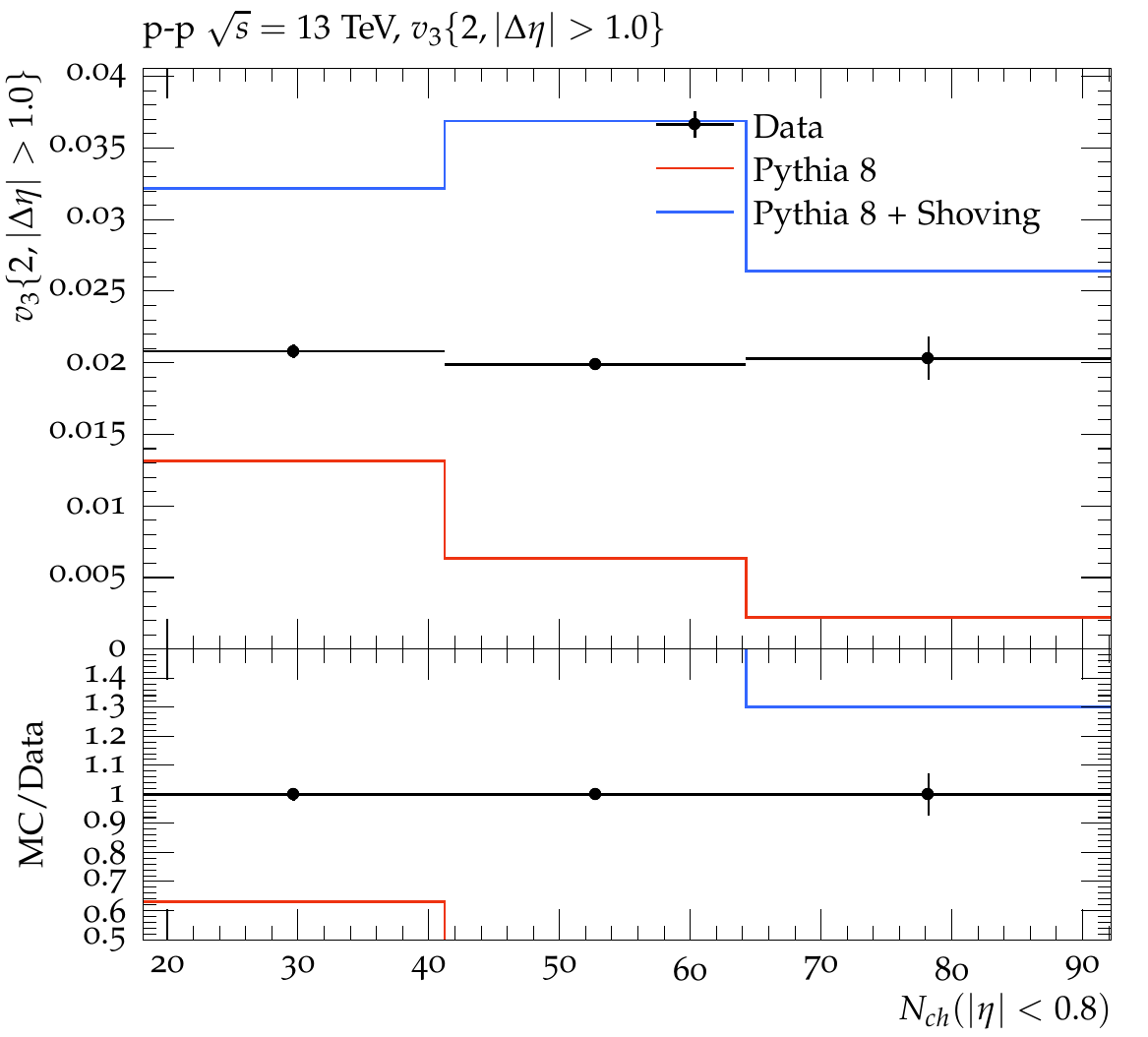}
	    \includegraphics[width=0.45\textwidth]{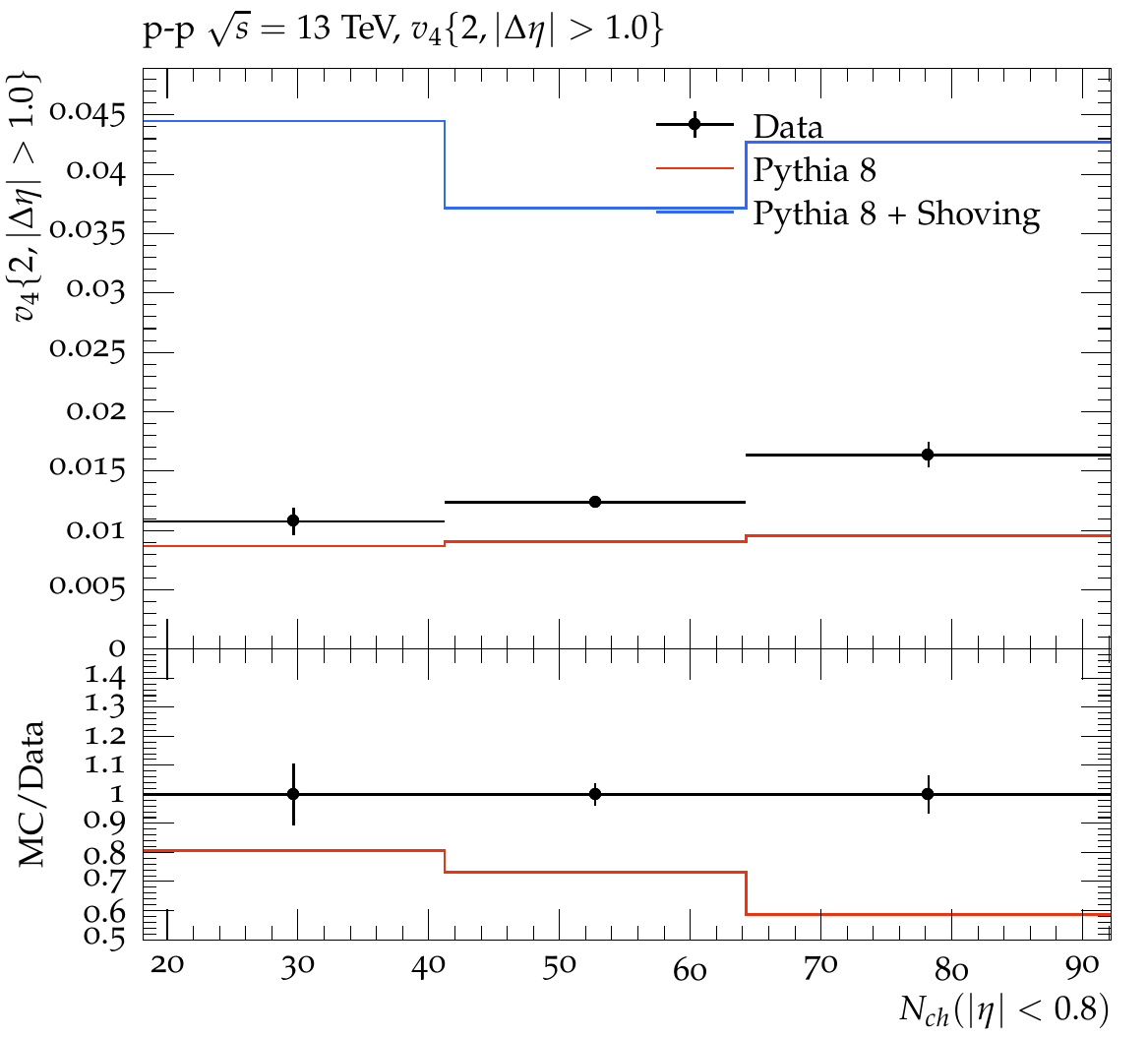}
    \end{center}
  \end{minipage}
  \caption{Comparison to $v_3\{2\}$ (left) and $v_4\{2\}$ (right) as function of multiplicity with ALICE high multiplicity trigger. Data from \pp collisions at $\sqrt{s} = 13$~TeV by ALICE \cite{Acharya:2019vdf}.}
  \label{fig:v3v4-pp}
}

In \figref{fig:v3v4-pp} comparisons to $v_3\{2\}$ (left) and $v_4\{2\}$ (right) are performed. While it is clear that shoving adds a sizeable contribution to both, it is equally clear that data is not very well reproduced. We remind the reader that any $v_n$ produced by the shoving model comes about as a response to the initial geometry, and the initial geometry used by default in \pythia, consists of two overlapping 2D Gaussian distributions. It was shown in \citeref{Bierlich:2019wld}, that applying more realistic initial conditions, can drastically change the eccentricities of the initial state in \pp collisions. So while the description at this point is not perfect, the observations that a clear effect is present, bears promise for future studies. Further on, correlations between flow coefficients, the so-called symmetric cumulants \cite{Bilandzic:2013kga,Albacete:2017ajt},
will be an obvious step. But at this point, without satisfactory description of the $v_n$'s themselves, it is not fruitful to go on to even more advanced observables.

\FIGURE[t]{
  \centering
  \begin{minipage}{1.0\linewidth}
    \begin{center}
	    \includegraphics[width=0.45\textwidth]{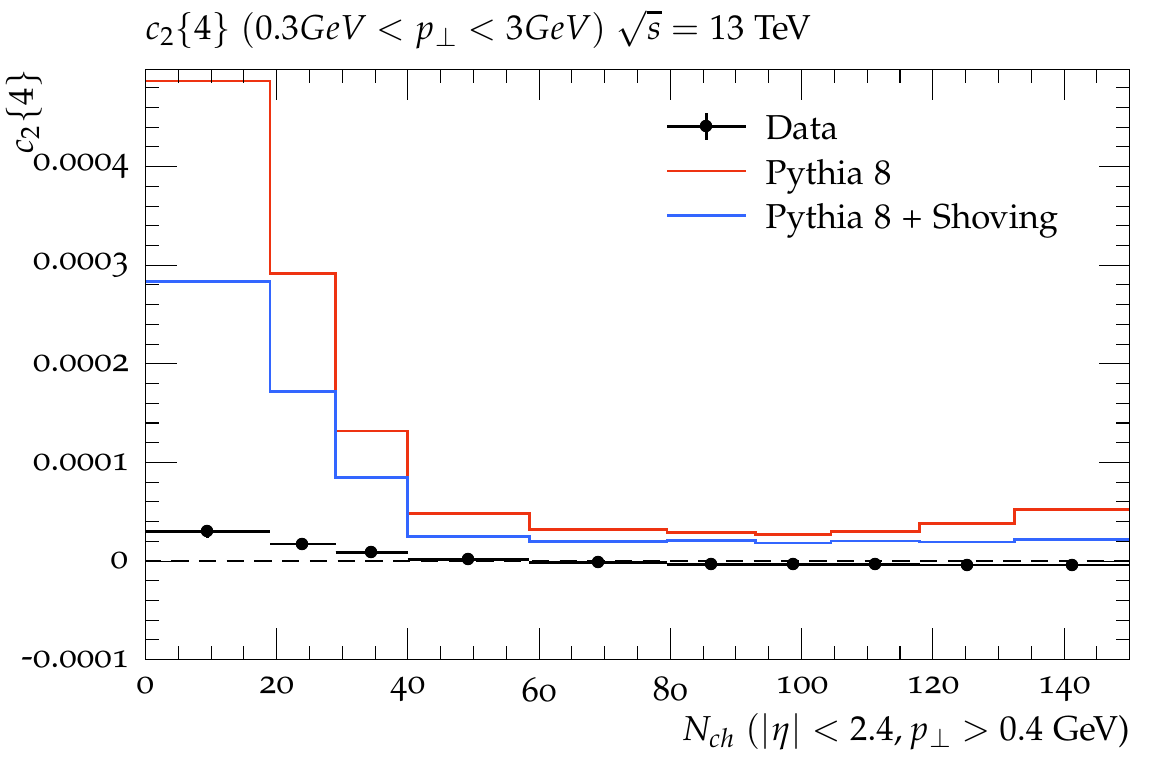}
	    \includegraphics[width=0.45\textwidth]{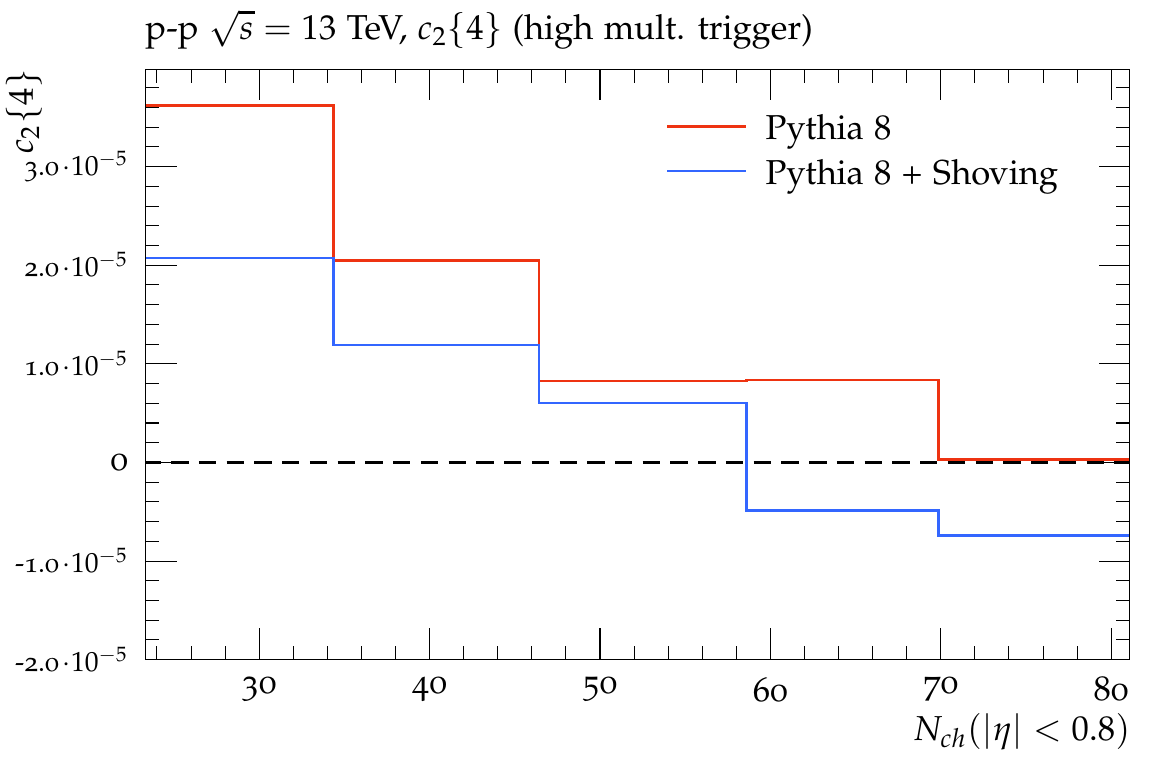}
    \end{center}
  \end{minipage}

  \caption{The four-particle cumulant $c_2\{4\}$, compared to data from CMS \cite{Khachatryan:2016txc} (left), and with the ALICE high multiplicity trigger (no data, right).}
  \label{fig:c24-pp}
}

Finally, in \figref{fig:c24-pp}, we show results for the four-particle cumulant $c_2\{4\}$. We briefly remind the reader about some definitions. The 2- and 4-particle correlations in a single event are given by the moments  \cite{Bilandzic:2010jr}:
\begin{eqnarray}
\langle 2\rangle &=& \langle \sum_{i,j} \exp[in(\phi_i - \phi_j)] \rangle \nonumber \\
\langle 4\rangle &=& \langle \sum_{i,j,k,l} \exp[in(\phi_i + \phi_j-\phi_k-\phi_l)] \rangle.
\label{eq:azimutmoments}
\end{eqnarray}
The averages are here taken over all combinations of 2 or 4 non-equal particles in one event.
The four particle cumulant is the all-event averaged 4-particle azimuthal correlations, with the 2-particle contribution subtracted:
\begin{equation}
    \label{eq:c24}
    c_n\{4\} = \langle \langle 4 \rangle \rangle - 2 \langle \langle 2 \rangle \rangle^2.
\end{equation}
The double average here means first an average over particles in one event, and then average over all events.

As discussed in \citeref{Borghini:2001vi}, when the correlation is dominated by flow and the multiplicity is high, then
the flow coefficient $v_2\{4\}$ is given by $\sqrt[4]{-c_2\{4\}}$. Clearly $c_2\{4\}$ must be \textit{negative} for this to be realized. This, in turn, means that the relative difference between the 2 -and 4-particle azimuthal correlations, must be right from \eqref{eq:c24}. As it was also pointed out in \citeref{Borghini:2001vi}, the non-flow contribution to four-particle correlations is much smaller than for two-particle correlations, as the cumulant becomes flow-dominated when $v_n \gg 1/M^{3/4}$ ($M$ is the multiplicity) in the former case, but only when $v_n \gg 1/\sqrt{M}$ in the latter. In a \pp collision $M$ is small compared to a heavy ion collision, and it can therefore be reasonably expected that the four particle correlations will only be flow dominated at sufficiently high multiplicity. Since data show a real $v_2\{4\}$, the importance of the sign of $c_2\{4\}$ in model calculations for \pp, have recently been highlighted \cite{Zhao:2017rgg,Adolfsson:2020dhm}. Importantly, standard hydrodynamic treatments do not obtain a negative sign of $c_2\{4\}$ in \pp collisions, even with specifically engineered initial conditions \cite{Zhao:2020pty}.

In the results from the shoving model in \figref{fig:c24-pp}, we note that while a negative $c_2\{4\}$ is not produced when comparing to CMS results, it is produced in high multiplicity events in the ALICE acceptance, using the high multiplicity trigger. There are several possible reasons for this apparent discrepancy. The acceptances are quite different, and since the sign of $c_2\{4\}$ is an observed characteristic, rather than a fundamental feature of the model, it is difficult to point out why a given model should produce different results in different acceptances -- though it is possible. More interesting, is the possible effect of the high multiplicity trigger. In \figref{fig:c24-pp} (left), it is seen that both default \pythia, as well as \pythia with the shoving model, over-predicts $c_2\{4\}$ at low-multiplicity by a large margin. As noted in the original paper, this is also the case for the 2-particle cumulant. A reasonable explanation for this over-prediction could be, that \pythia collects too many particles in mini-jets in general. With a high multiplicity forward trigger, a strong bias against this effect is put in place,
and the underlying model behaves more reasonable. In any case, the finding of a negative $c_2\{4\}$ in high-multiplicity events with the shoving model is an interesting and non-trivial result, which will be followed up in a future study.

\subsection{Results in Pb-Pb collisions}
\label{sec:aa-results}

We now turn to Pb-Pb collisions, where we use the \angantyr model in PYTHIA keeping the same settings as in section \ref{sec:pp-results}. The results for Pb-Pb collision events at 5.02 TeV has been compared to ALICE data points via \rivet routines.\footnote{The \rivet routines are not yet validated by the experimental community.} The anisotropic flow coefficients plotted here have been calculated, as in the previous section, using multi-particle cumulant methods as done in the ALICE experiment.
 
Centrality measures used in these analyses are of two kinds: for the plot in \figref{fig:Pb-multiplicity} we use the centrality binning of the generated impact parameter by \angantyr, and for the plots in \figref{fig:Pb-v2}, we use \angantyr generated centrality binning which mimics the experimental centrality definition where in ALICE the binning is in the integrated signal in their forward and backward scintillators. However, the difference between the two centrality measures is small in Pb-Pb collisions \cite{Bierlich:2020wms}.

\FIGURE[t]{
	\centering
	\includegraphics[width=0.6\textwidth]{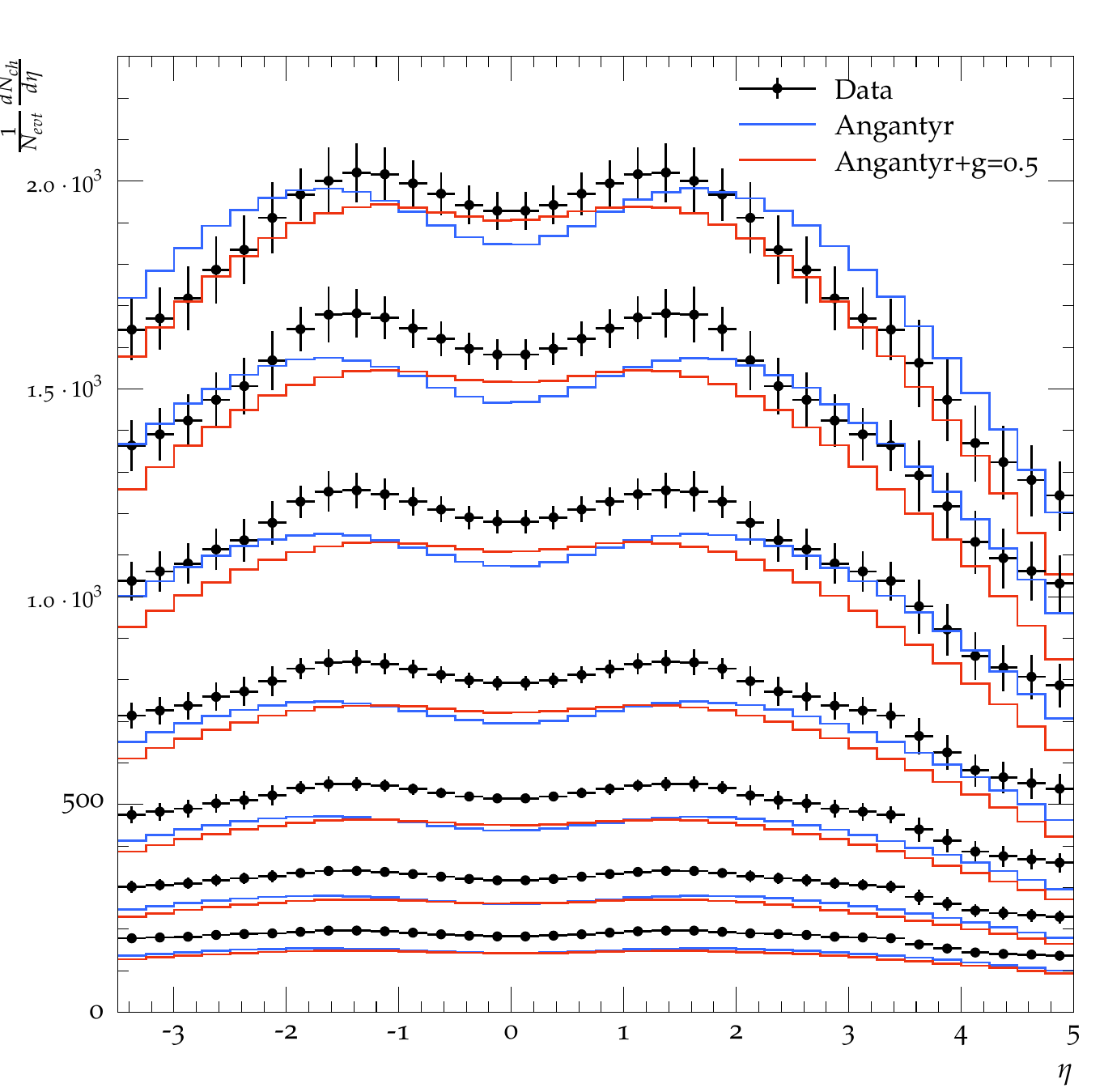}
	\caption{Charged particle multiplicity over a wide range of $\eta$ in Pb-Pb collisions at $\sqrt{s_{NN}} =5.02$~TeV for centralities 0-5\%, 5-10\%, 10-20\%, ..., 50-60\% \cite{Adam:2016ddh}.}
	\label{fig:Pb-multiplicity}
}

In \figref{fig:Pb-multiplicity}, we plot the charged particle multiplicity for seven centrality classes (0-5\%, 5-10\%, 10-20\%, 20-30\%, 30-40\%, 40-50\%, 50-60\%) as a function of pseudorapidity in the range $-3.5 < \eta < 5$ for $\sqrt{s_{NN}}=5.02$ TeV in Pb-Pb collisions comparing it to the study performed by ALICE\cite{Adam:2016ddh}. We use this figure as our control plot to check that when we turn shoving on, the description of other observables are not destroyed. 

We observe that this implementation fairly well preserves the \angantyr
description of the multiplicity distributions. The overall
multiplicity of the shoving curve is however a bit lower when compared
to default Angantyr, which is because of the increased \texttt{pT0Ref}
as mentioned in \ref{sec:pp-results}.  Also, as discussed in
\sectref{sec:generating-shoving}, when strings are shoved and the
particles on average get a larger $p_\perp$ which also means that they
come closer together in pseudorapidity. The overall effect is that
particles are generally \textit{dragged} closer towards mid-rapidity,
reducing the two-humped structure seen for plain \angantyr.

We will look into further improvement of the multiplicity description by shoving through tuning, normalization of the distribution functions and accurate description of centrality as in experiments in the future.

\FIGURE[t]{
	\centering
	\begin{minipage}{1.0\linewidth}
		\begin{center}
			\includegraphics[width=0.42\textwidth]{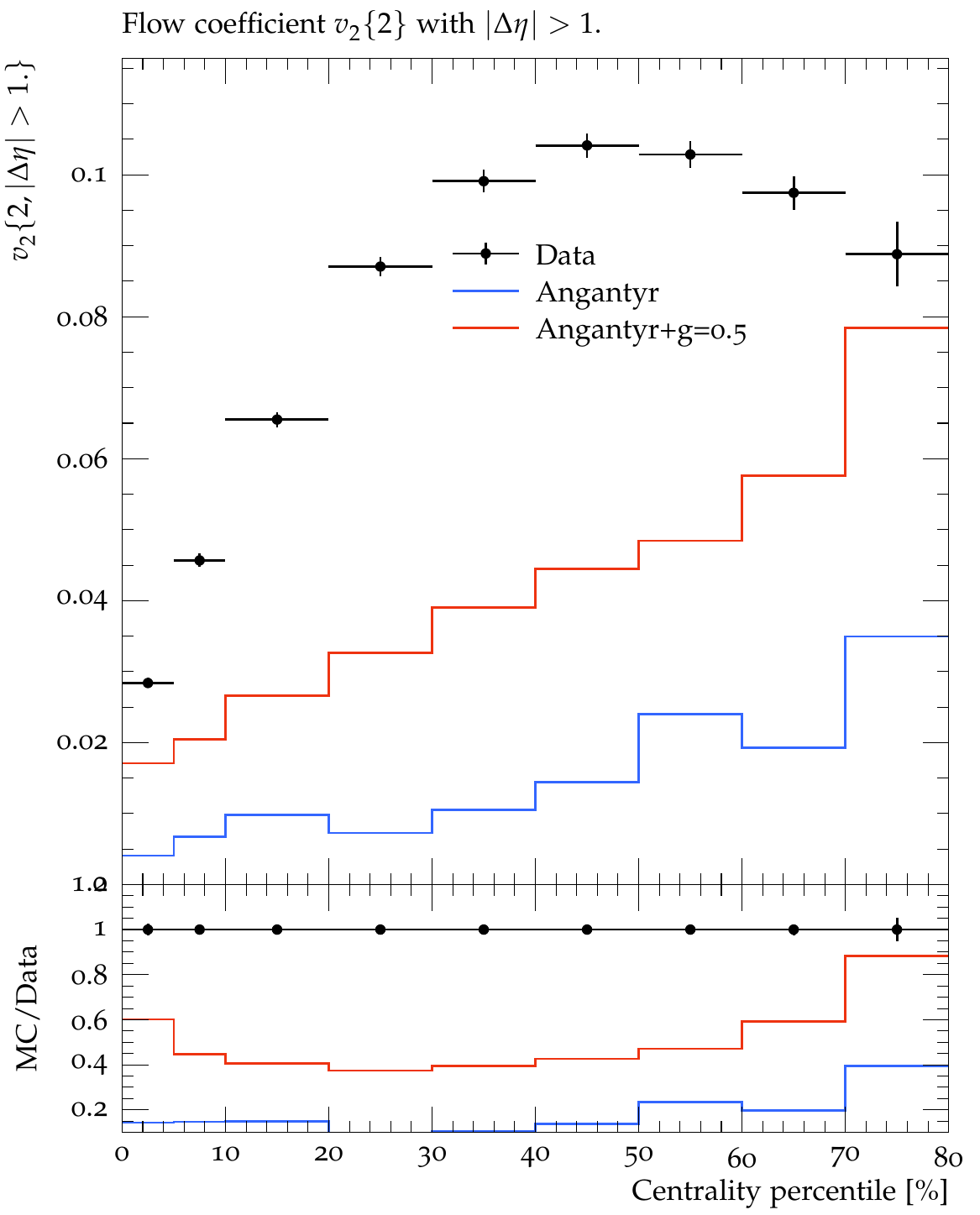}
			\includegraphics[width=0.42\textwidth]{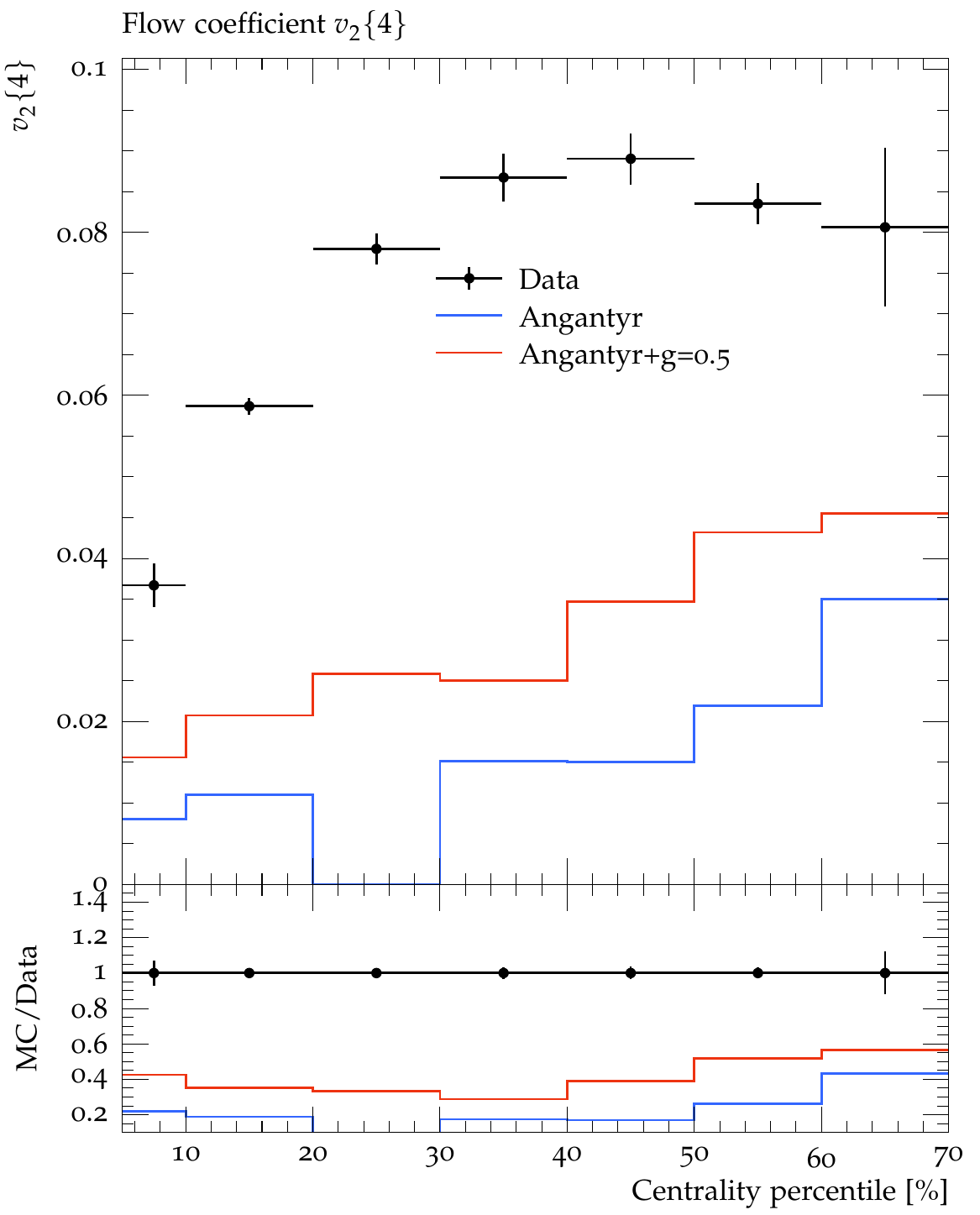}
		\end{center}
	\end{minipage}
	\caption{The flow coefficient $v_2\{2\}$(left) with $|\Delta \eta| > 1 $ and $v_2\{4\}$(right) for $0.2 < p_\perp < 5.0$~GeV  in Pb-Pb collisions at $\sqrt{s_\mathrm{NN}} =5.02 $~TeV\cite{Adam:2016izf}.}
	\label{fig:Pb-v2}
}

\Figref{fig:Pb-v2} presents the centrality dependence of the harmonic flow coefficient $v_2$ from two-particle cumulant on the left with $|\Delta \eta| >1$ and four-particle cumulants on the right, integrated over the $p_\perp$ range $0.2 < p_\perp < 5.0$~GeV for 5.02 TeV Pb-Pb collisions \cite{Adam:2016izf} for generated centrality. We note that \angantyr with shoving results in an increased $v_2$ in the right direction with respect to data. We see in data that $v_2\{2\}$ increases from central to peripheral collisions, reaching a maximum of around 0.10 between $40-50$\% centrality. $v_2\{4\}$ also shows a similar behaviour with a maximum around 0.09 between $40-50$\% centrality.\footnote{In the plot for $v_2\{4\}$, there is lack of statistics in the centrality bin 20-30\% for \angantyr without shoving.} String shoving result clearly lacks the curvature of the data points, but doubles the contribution in $v_2\{2\}$ as compared to \angantyr. The underlying cause for this behaviour is that the current implementation of shoving alone is not sufficient to generate the large overall response to the anisotropy in the initial collision geometry of the nuclei. An increased $g$ factor or delayed hadronization time or an early onset of shoving, and a combination of these factors, do not give rise to enough $v_2$ either. 

In \sectref{sec:pbtoy}, we showed that shoving can generate sufficient
$v_2$ as seen in data with completely straight strings without any
gluon kinks. With \angantyr, we have more realistic final states with
many and often soft gluon emissions from the multi-parton interactions
and the initial- and final-state evolution models, which hinder the
process of strings shoving each other by cutting short their
interaction time, hence resulting in the overall observation of the
lack of enough transverse \pushes generated via this mechanism.

\section{Conclusions and outlook}
\label{sec:conclusions}

In this paper we have  argued that a hot thermalized plasma is
not necessarily formed even in central \AA\ collisions, not even in
Pb--Pb collisions at the highest attainable energies at the
LHC. Instead we note that the string-based approach to simulating
hadronic final states in the \angantyr model in \pytppp gives a very
reasonable description of the number and general distribution of
particles in \AA\ events, and take this as an incentive to study
string hadronization in dense collision systems more carefully.

Our string picture is qualitatively different from the more conventional picture, where
the colliding nuclei are described in terms of a CGC, that in the moment of collision 
turns into a so-called \textit{glasma}, which very soon decays into a thermalized QGP. 
Similar to the string picture, the glasma has longitudinal fields stretched between the 
nucleus remnants, and these fields are kept together in flux tubes as the remnants move 
apart. There are, however, also essential differences. The glasma turns rapidly into a thermalized 
plasma. Such a plasma expands longitudinally in a boost invariant way, with decreasing energy density as a result. The initial density must  therefore be quite high to give the observed particle density after freezeout. In contrast the energy density in the strings is constant up to the time for hadronization. When the strings become longer, the energy in the new string pieces is taken from the removing nucleus remnants, and not from depleting the energy in the strings already formed.
In the string scenario we estimate the energy density at mid-rapidity to around 5~GeV/fm$^3$
(in \PbPb\ at the LHC), while in the glasma we find that it ought to be one or two orders of magnitude higher. 

The low energy density in the string scenario implies that the vacuum condensate is very important to form the strings.
The break-down of the glasma is often motivated by the so-called
``Nielsen--Olesen instabilities''. These authors showed that a longitudinal chromo-electric field added to the QCD Fock vacuum is instable, and transverse fields grow exponentially. 
This growth does not go on forever. Instead higher order
corrections will lead to a Higgs potential analogous to the potential describing the 
condensate of Cooper-pairs in a superconductor or the vacuum condensate in the EW Higgs model. Adding a (not too strong) linear field to this non-trivial ground state will then give flux tubes, similar to the vortex lines in a superconductor.
We take this as an indication that it may indeed be
possible that strings can be formed and actually survive the initial
phase of the collision, without a thermalized plasma being formed.
(The energy density needed in the glasma may, however, be strong enough to destroy
the "superconducting" phase.)

Another important difference between the two scenarios is that the glasma contains both chromo-electric and chromo-magnetic fields, not only chromo-electric fields as in the string picture. This implies CP-violating effects in the glasma, but this feature is not discussed in this paper.

In an earlier paper we argued that in a dense environment where the
strings overlap in space--time, they should repel each other and we
showed with a very simple model that this could induce flow in
\pp. Here we have motivated the model further, and compared to lattice
calculations to estimate the transverse shape and energy distribution
of the string-like field. We have also improved the implementation of
the model, where the strings no longer have to be completely parallel
in order to calculate the force between them. Instead we show that for
a pair of arbitrary string pieces, we can always find a Lorentz frame
where they will be stretched out in parallel planes, allowing us to
easily calculate the force there. We have further improved the time
discretization, and instead of processing string interactions in fixed
time intervals, the shoving model is implemented as a parton shower
with dynamical time steps, greatly improving the computational
performance of the model. Finally we have improved the procedure for
transferring the \pushes from string interactions to final state
hadrons.

The implementation used for obtaining the results presented here is
not yet quite complete. Although it circumvents the production of huge
amounts of soft gluons in the shoving, which was a major problem for
our previous implementation, it still has a problem with dealing with
the soft gluons that are already there from the initial- and
final-state parton showers. Gluon kinks loose energy with twice the force 
compared to a quark. A soft gluon therefore will soon loose all its energy, and new
straight sections of the string will be formed, which in the current
implementation are not taken into account.

In addition the implementation only allows shoving at points in 
space--time after the strings have expanded to the equilibrium size,
$R_S\sim\tauS$, and before they start to break up in hadrons at proper times
around \tauH. Clearly shoving should be present also at times before
\tauS, and the force between very close strings should then be higher, but
our implementation currently cannot handle situations with varying
string radii. Also, at times later than \tauH, after the string
break-up, one could expect some shoving between the hadrons being
formed, and after they are formed one needs to consider final state
re-scattering.

In light of these shortcomings of the current implementation, it is not
surprising that when we apply the model to complete partonic final
states generated by \angantyr, we cannot quantitatively reproduce the
amount of $v_2$ measured in experiments. But we do see that the
shoving actually does give rise to flow effects. We also see that in
\pp\ we currently get a bit too large $v_2$, but we would like to
emphasise that we have here not tried to do any tuning of the model
parameters. Instead these are kept at canonical values,

To investigate the model further and to make it plausible that the
shoving model, when implemented in full, actually may be able to
reproduce also quantitatively the flow effects measured in \AA\
collisions, we looked at what happens when applying it to a toy model
of the initial state. To account for the missing string pieces
due to soft gluons, we here
used parallel strings without gluon
kinks. The strings are randomly distributed with variable density in an ellipsoid shape
in impact parameter space. In this way the shoving was unhampered by
soft gluons, and we found that the model is able to get the azimuthal
anisotropies in momentum space as expected from the eccentricity of
the shape in impact parameter. By then matching the string density to
the one we have in \angantyr for different centralities, we could also
see that the resulting $v_2$ was much closer to measured data.

Showing that we can get reasonable azimuthal anisotropies in \AA\
collisions using a purely string-based scenario is, of course, not
enough to prove that a thermalized QGP is \textit{not} formed in such
collisions. To do this we need to also be able to describe other
measurements, such as strangeness enhancement and jet quenching and, more
importantly find new observables where a string based scenario
predicts results that cannot be reconciled with the QGP picture. And for
this we not only need to improve the implementation of the shoving
model, but also revisit our rope model and also our ``swing'' model
for colour reconnections. The rope model has been shown to give
reasonable descriptions of strangeness enhancement in high
multiplicity \pp, and using the parallel frame presented here, we
should be able to get a better handle on the space--time picture of
the string overlaps needed to be able to apply it in \AA. Also for the
swing model we can take advantage of the parallel frame to properly
understand which partons may reconnect and when and where they may do
so. And since in the parallel frame hard and soft partons are treated on an equal
footing this could also have interesting effect on jets.

In the end we hope that these models will be implemented in \pytppp
together with the \angantyr model, so that we get a complete platform
for generating fully hadronic final states that can be compared to any
type of measurement in any kind of collision (\AA, \pA, \pp,
\ldots). This would then give us a perfect laboratory to investigate a
purely string-based picture as an alternative to the conventional QGP
approach.

\section*{Acknowledgments}

This work was funded in part by the Knut and Alice Wallenberg
foundation, contract number 2017.0036, Swedish Research Council, contracts
number 2016-03291, 2016-05996 and 2017-0034, in part by the European
Research Council (ERC) under the European Union’s Horizon 2020
research and innovation programme, grant agreement No 668679, and in
part by the MCnetITN3 H2020 Marie Curie Initial Training Network,
contract 722104.

\appendix

\section{Vortex lines in a superconductor}
\label{sec:vortex}

The microscopic properties in a superconductor, the magnetic field, $H$, and the
condensate wavefunction (order parameter), $\psi$, can be determined by
the LG equations (see \eg\ \citeref{ref:deGennes}). The
theory contains two characteristic lengths, the penetration depth $\lambda$
for the magnetic field and the coherence length $\xi$ for the condensate. 
In its generalization to a relativistic theory,
the Lagrangian contains a Higgs potential for the condensate. Here the
lengths $\lambda$ and $\xi$ correspond to the (inverse) masses of the gauge
boson and the Higgs particle respectively. 

For a flux tube the wavefunction $\psi$ is undetermined along a ``vortex line'',
and has a phase changing by $2\pi n$ when going around the vortex line, with
$n$ an integer. The total flux in the flux tube is then quantized to $n$ times
a flux quantum $\Phi_0= 2\pi/q$, where for a normal superconductor $q=2e$ is 
the charge of a Cooper pair. This quantum would also correspond to the charge of a
magnetic monopole. The change in phase is related to a vortex-like current in the
condensate, which keeps the flux confined within the flux tube.

At the boundary between a normal and a superconducting phase, the pressure
from the condensate and the magnetic field balance each other. The condensate
goes to zero over a distance $\xi$ in the superconductor, and the magnetic
field is suppressed over a distance $\lambda$. As a result, when $\xi$ is
larger than $\lambda$ (or more exactly $ \xi>\sqrt{2}\,\lambda$), both the
condensate and the field are suppressed over a range $\xi-\lambda$. This is a
\emph{type I} superconductor, and it implies that the surface provides a
positive contribution to the energy. In equilibrium the surface then tends to
be as small as possible. If in contrast $\lambda$ is larger than $\xi$
(\emph{type II} superconductor), the condensate and the field can coexist over
a range $\sim\lambda-\xi$, and the surface provides a negative contribution to
the energy, favouring a large surface. If the flux tube has more than one flux
quantum, there is then a 
tendency to split it into a number of vortices, each with one unit of flux.
In case of a large total flux, there is a repulsive force between nearby
flux tubes. The system will then tend to expand, forming an ``Abrikosov lattice''.

The interaction between the condensate and the electromagnetic field in a
superconductor is described by the LG equations, which in its
relativistic generalization corresponds to the Abelian Higgs model relevant for 
the Abelian projection of the QCD field. The Lagrange density is here given by
\begin{equation}
\mathcal{L} = -\frac{1}{4} F_{\mu\nu}F^{\mu\nu} +
[(\partial_\mu +ieA_\mu)\psi^*][(\partial_\mu -ieA_\mu)\psi] 
-\alpha |\psi|^2 -\frac{\beta}{2} |\psi|^4.
\label{eq:LGeq}
\end{equation}
When the parameter $\alpha$ is smaller than zero, the scalar field $\psi$ forms a 
condensate $\psi = \psi_0= \sqrt{-\alpha/\beta}$. The mass of the Higgs particle
and the massive gauge boson, given by $\sqrt{-2\alpha}$ and $e\sqrt{-2\alpha/\beta}$
respectively, correspond to (the inverse of) the coherence length $\xi$ and the 
penetration depth $\lambda$.
The LG equations are obtained from Euler-Lagrange's equations varying 
$\psi$ (or $\psi^*$) and $A_\mu$.

In an extreme type II superconductor with $\xi \ll \lambda$, the
LG equations have a solution $\psi = const. \,\,e^{-i\phi}$ (for  $\rho > \xi$),
for a vortex line which carries one unit of flux. The corresponding magnetic field is here given by
\begin{equation}
H (\rho)= C\cdot K_0(\rho/\lambda),
\label{eq:Clem1}
\end{equation}
where $C=\Phi/(2\pi \lambda^2$) is a constant, $\Phi$ is the total flux, and
$K_0$ is a modified Bessel function. The Bessel function has a logarithmic
dependence on $\rho$ for $\rho<\lambda$, but falls exponentially for 
$\rho>\lambda$. The field is confined within this range by an electric current
$j=\lambda \,|\mathrm{curl} \,\mathbf{H} | = (C/\lambda)\, K_1(\rho/\lambda)$ (also valid for
$\rho>\xi$). In 
this extreme case, when $\xi$ is very small, the contribution to the energy from
destroying the condensate is also small, and the energy of the flux tube, the string tension
$\kappaS$, is given by the sum of the field energy and the energy in the current:
\begin{equation}
\kappaS=\int_\xi d^2\! \rho\, \frac{1}{2}\left\{ H(\rho)^2 + \lambda^2
  (\mathrm{curl}\, \mathbf{H}(\rho))^2\right\} .
\label{eq:Clem2}
\end{equation}
We note that the energy is dominated by the contribution from the current, where
$K_1(\rho/\lambda)$ is singular and $\sim \lambda/\rho$ for small $\rho$.
Thus the total energy is proportional to $\ln(\lambda/\xi)$ for very small $\xi$.

When $\xi$ is smaller than $\lambda$, but
not close to zero, the following approximate solution is given by Clem \cite{ref:Clem}:
\begin{equation}
 H(\rho)= C \,K_0(x_\perp/R), \,\,\,\,\, \mathrm{with} \,\,\,R=  \sqrt{\rho^2 +
   \xi^2_v}\,\,\,\, \mathrm{and}\,\,\, C =\Phi/(2\pi\lambda\, \xi_v\,
   K_1(\xi_v/\lambda)). 
   \label{eq:clem}
\end{equation}
The new distance scale $\xi_v$ depends on the ratio $\kappaLG\equiv \lambda/\xi$, and
is close to $\xi$ for $\kappaLG \approx 1/\sqrt{2}$, \ie\ for a
superconductor on the border between type I and type II. The expression in \eqref{eq:clem}
satisfies the one of the LG equations (obtained by varying the field $A_\mu$) and 
Amp\`er's equation $\mathbf{j}= 
\mathrm{curl}\, \mathbf{B} = \mathrm{curl}\,\mathrm{curl}\, \mathbf{A}$.
It does, however, not satisfy the equation obtained by varying $\psi$, and for $\xi>\lambda$ this equation can be badly violated.

In a superconductor there are magnetic flux tubes, and magnetic monopoles would be confined. 
In QCD colour-electric  flux tubes and colour-electric charges are confined. Thus for the 
Abelian projection the fields $F_{\mu\nu}$ will be replaced by the dual fields 
$\widetilde{F}_{\mu\nu}$, as discussed in \sectref{sec:condensate}.

\bibliographystyle{JHEP}
\bibliography{shoving-ref}

\end{document}